\documentclass[]{aa}  
\usepackage{graphicx}
\pdfoutput=1

\usepackage{txfonts}
\usepackage{natbib}
\bibpunct{(}{)}{;}{a}{}{,} 

\def\thetadef{15$^\circ\,$}
\def\thetadefweights{16$^\circ\,$}

\def\bval{0.67}
\def\bvalang{$39^\circ\,$}
\def\bvalerrproc{1\%}
\def\bvalangweights{$26^\circ\,$} 
\def\bvalerrprocweights{3\%}
\def\sep1{$0^\circ\,$}
\def\sep2{$34^\circ\,$}
\def\sep3{$46^\circ\,$}

\begin{document}

   \title{Mathematical properties of the \emph{SimpleX\/} algorithm}

   \author{C. J. H. Kruip,
          \inst{1}
          J.-P. Paardekooper
          \inst{1}
          B. J. F. Clauwens,
          \&
          V. Icke
          \inst{1}\fnmsep\thanks{OK Jelle, this one's for you}
          }

   \offprints{C. J. H. Kruip}

   \institute{\inst{1}Sterrewacht Leiden, Postbus 9513, 2300RA Leiden, The Netherlands\\
              \email{kruip@strw.leidenuniv.nl}}

   \date{Received *** 2009; accepted ***}

 
  \abstract
  {Analytical and numerical analysis of the \emph{SimpleX\/} radiative transfer
    algorithm, which features transport on a Delaunay triangulation.}
  {Verify whether the \emph{SimpleX\/} radiative transfer algorithm conforms to
    mathematical expectations, to develop error analysis and present
    improvements upon earlier versions of the code.}
  {Voronoi-Delaunay tessellation, classical Markov theory.}
  {Quantitative description of the error properties of the \emph{SimpleX\/}  method.
    Numerical validation of the method and verification of the analytical results. 
    Improvements in accuracy and speed of the method.}
  {It is possible to transport particles such as photons in a
    physically correct manner with the \emph{SimpleX\/} algorithm. This
    requires the use of weighting schemes or the modification of the point process underlying the transport grid. We have explored and applied several
    possibilities.}


\maketitle


\section{Introduction}\label{sec:introduction}

At present one of the major challenges in computational astrophysics is to correctly account for radiative transfer in realistic macroscopic simulations. Whether one considers the formation of single stars or the evolution of merging galaxies, incorporation of radiative transfer is the next essential step of physical realism needed for a deeper understanding of the underlying mechanisms.

With the advent of the computer as an important catalyst, the last century has seen a myriad of efforts to solve the equations of radiative transport within a numerical framework. The resulting algorithms cover a wide range of applications where generally a method is tailored to a specific problem. In most cases, specialisation either means the choice for a specific physical scale (consider detailed models of stellar atmospheres or large-scale cosmological simulations) or emphasis on physical processes relevant for the problem at hand.

\subsection{The \emph{SimpleX\/} algorithm}

The \emph{SimpleX\/} algorithm for radiative transfer, which is the subject of this text, has been designed to steer clear from these limitations. Due to a modular structure, different physical processes can be included straightforwardly allowing different areas of application. More importantly, the method has no inherent reference to physical scale and thus can be applied to problems with typical length scales that lie many orders of magnitude apart. Moreover, due to its adaptive nature, many orders of magnitude in optical depth and spatial resolution can be resolved in the same simulation.

Conceived by \citet{2006PhRvE..74b6704R} and implemented by \citet{2007PhDT1R}, the \emph{SimpleX\/} algorithm solves the general equations of
particle transport by expressing them as a walk on a graph. In this sense the method can be thought of as a Markov Chain on a closed graph. Transported quantities travel from node to node on the graph, where each transition has a given probability.

This approach has several advantages which we would like to highlight here. More specifically \emph{SimpleX\/} 

\begin{itemize}
\item does not increase its computational effort or memory use with the number of sources in a simulation and consequently treats for instance scattering by dust and diffuse recombination radiation without added computational effort
\item naturally adapts its resolution to capture the relevant physical scales (e.g. expressed in photon mean free path lengths)
\item works in parallel on distributed memory machines
\item is compatible with grid-based as well as point based hydrodynamics codes (where the latter is the more natural combination due to the point-based nature of both  \emph{SimpleX\/} and SPH)
\item is computationally cheap due to the local nature of the Delaunay grid
\end{itemize}

\subsection{Error analysis}

Covering many orders of magnitude in spatial resolution (and optical depth) is a considerable challenge for radiative transfer methods in general. Often, approximations need to be employed to keep the problem tractable, in turn making a robust error analysis difficult.  

For the vast majority of numerical methods the errors are measured by comparison with a fiducial run of the code. This is usually a simulation wherein many more time steps and/or a higher spatial resolution is used than would be normally feasible. Usually, some sort of convergence to a solution is observed and this solution is accepted as the correct one, at least within the possibilities of the method.

For \emph{SimpleX\/} we cannot do such a convergence test in general\footnote{Although this is possible if the direction conserving transport to be introduced in Sec.~\ref{sec:DCT} is used.  Alternatively, several instances of the same simulation with a different random seed for the grid construction can be averaged to obtain error estimates as well.}. When the spatial resolution is increased far beyond the resolution dictated by the local mean free path length of the photons, several effects to be described in Sec.~\ref{sec:anis_its_cons} will tend to make the radiation field more diffuse, \emph{decreasing} instead of \emph{increasing} accuracy. This property of \emph{SimpleX\/} is not a weakness but instead inherent to the fact that the method uses a \emph{physical} grid wherein deviations from its natural resolution often imply deterioration of the solution. 

A great advantage is that the \emph{SimpleX\/} algorithm, due to its mathematically transparent nature,  allows for an analytical assessment of its error properties. The resulting prescriptions are quite general and can be applied to different regions in parameter space, an advantage over the `numerical converge approach' usually applied in the error analysis of radiative transfer methods. The development of analytical descriptions and discussion of their implications will be the main focus of this text. 

\subsection{Outline}

Given the desirable properties of the \emph{SimpleX\/} algorithm, we aim to assess if the various inaccuracies that arise inevitably in a numerical method can counterweigh  \emph{SimpleX\/}s virtues in any case. 

We will show that the use of high dynamic range Delaunay triangulations as the basis for radiative transfer introduces systematic errors which reveal themselves as four distinct effects: diffusive drift, diffusive clustering, ballistic deflection and ballistic decollimation. 

We will describe and quantify these effects and, subsequently, provide suitable solutions and strategies. 

Following this, we will demonstrate how the derived measures can be used to constrain the translation of a physical problem to a transport graph in such a way as to avoid or minimise errors. 

Finally, we will discuss the results in a broader picture and outline related and future work; specifically the recent implementation of a dynamically updating grid in \emph{SimpleX\/}.

Although developed in the context of \emph{SimpleX\/}, many of our results are relevant for other transport algorithms that work on non-uniform grids (e.g. AMR grids or SPH particle sets.)

\section{Radiative transfer on unstructured grids}

In this section we describe the three means of transport that constitute the \emph{SimpleX\/} algorithm. Starting from a general description of the equation of radiative transfer in inhomogeneous media, we will delve into the specifics of \emph{SimpleX\/}. This in turn asks for an introduction to the Delaunay triangulation which lies at the heart of our method and plays a central role in this text.

\subsection{General radiation transfer}

The equation of radiative transfer for a medium whose properties can change both in space and time is given by
\begin{equation}
\frac{1}{c}\frac{\partial I_{\nu}}{\partial t} + \hat{n}\cdot \nabla I_{\nu} = \eta_{\nu} - \chi_{\nu} I_{\nu}.
\end{equation}
Here $I_{\nu}\equiv I (t, \vec{x}, \mathbf{\Omega}, \nu)$ is the monochromatic specific intensity, $\hat{n}$ is a unit vector along the ray and $\eta_{\nu} $ and $\chi_{\nu} $ are the monochromatic emission and opacity coefficients respectively. 

If $\frac{1}{c}\frac{\partial I_{\nu}}{\partial t} \ll 1$ (in other words: if $I_{\nu}$ is not explicitly time dependent \emph{or} the time and space discretisation is such that $c$ can be considered infinite) this equation simplifies to
\begin{equation}
\hat{n}\cdot \nabla I_{\nu} =\eta_{\nu} - \chi_{\nu} I_{\nu}.
\label{eq:RT}
\end{equation}
If we take the spatial derivative along the ray and divide by $\chi_{\nu}$, Eq.(\ref{eq:RT}) can be rewritten as
\begin{equation}
\frac{\partial I_{\nu} }{\partial \tau} =S_{\nu} -  I_{\nu}
\label{eq:RTtau}
\end{equation}
where we have defined the source function $S_{\nu}\equiv \eta_{\nu} /\chi_{\nu}$ and optical depth $d\tau \equiv \chi_{\nu} ds$ and $s$ parametrises the distance along the ray. Eq.(\ref{eq:RTtau}) can be solved numerically if $\eta_{\nu} $ and $\chi_{\nu} $ are known locally. What we mean by `local' depends on the type of discretisation of the volume. The form of Eq.(\ref{eq:RTtau}) suggests that the optical depth is the most natural variable to discretise. Doing this will in general result in cells of unequal volume and, hence, an irregular computational mesh. As mentioned in Sec.~\ref{sec:introduction} and explained in more detail in Sec.~\ref{sec:application} we generate a set of nuclei which reflects the underying optical depth field and connect these nuclei with a Delaunay triangulation.

\subsection{A natural scale}

The fundamental idea behind the \emph{SimpleX\/} algorithm is the fact that there exist a natural scale for the description of radiative processes: the photons local \emph{mean free path}
\begin{equation}
l_{\mathrm{mfp}} \equiv \frac{1}{\sigma n}.
\end{equation}
Here $\sigma$ is the total cross section and $n$ the number density of the particles responsible for the extinction.
As photons travel typically one mean free path before interacting with the medium, information on a much smaller scale does not necessarily yield a deeper understanding of the physical problem at hand.

Following this line of reasoning, the next step is to choose a computational mesh that inherently carries this natural scale. In other words: use an irregular grid which resolution adapts locally to the mean free path of the photons traveling over it.\footnote{ In our current implementation we use a single frequency bin and an effective cross section so the mean free path depends on the number density and the point-to-point distance only.}

\subsection{The grid}

This computational mesh or transport graph, is constructed by first defining a point process that represents the underlying (physical) problem and, second, by connecting these points according to a suitable prescription.

Specifically, the point process is defined by prescribing the
local point field density as a function of the scattering and
absorption properties of the matter through which the particles
propagate. The resulting points are connected by means of the unique \emph{Delaunay
triangulation}~\citep{Delone1934}. Thus, the points that carry the triangulation, which we
call {\it nuclei\/} in the remainder of this text, are connected in a
graph, along whose connecting lines (called \emph{edges}) the particles are required to travel. 

This choice for constructing a transport graph has several advantages: first, the connection with the physical processes is evident; second, the Delaunay triangulation (and its dual, the \emph{Voronoi tessellation}~\citep{Voronoi1908}) is unique and is in many respects the optimal unstructured partitioning of space \citep{Schaap:2000p855}; third, the Voronoi-Delaunay structures carry all relevant information needed for the transport process, making the transport step itself very efficient. 

We note that the use of Voronoi-Delaunay grids does not imply that the transfer of photons has to proceed along the edges of the Delaunay triangulation.
Another approach would be to trace rays through either the Voronoi cells or Delaunay simplices. This ray-tracing could be done classically with long-characteristics \citep[e.g][]{1984frh..book.....M} or using Monte Carlo methods \citep[e.g][]{Abbott:1985p1320}. As said, in SimpleX, the photons travel from cell to cell interacting with the medium as they go along. In this sense, the transport of radiation is treated as a local phenomenon and the global nature of the radiative transfer problem is dealt with by sufficient iterations of this local transport. The intimate relation between the structure of the grid and the transport of the photons is one of the reasons why the \emph{SimpleX\/} algorithm is exceptionally fast.

The construction of the grid itself is a task performed by dedicated software which is freely available on the web. As said, once given the generating nuclei, the Voronoi-Delaunay structure is unique. It is  in creating the distribution of generating nuclei that we can manipulate the properties of our computational grid. The translation from a given density (opacity) field either as particles or as cells to a Voronoi-Delaynay grid optimal for \emph{SimpleX\/} is a fundamental part of the algorithm.

\subsection{Voronoi-Delaunay structures}\label{sec:voro_delau}

 Because it plays a central role in this text, we now proceed with a concise introduction to the Voronoi-Delaunay grid.

\begin{figure}[!ht]
  \begin{center}
    \includegraphics[width=0.45\textwidth]{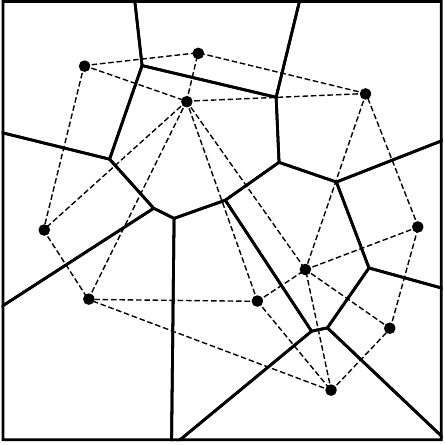}
    \caption{Voronoi tessellation of the plane (solid lines). 
      Each cell contains all points that are closer to its nucleus
      (indicated by a dot) than to any other nucleus. The corresponding Delaunay triangulation is shown in dashed lines. Note: only visible nuclei are included in
      the triangulation.}\label{fig:VoronoiDelaunay}
  \end{center}
\end{figure}

Let a set of nuclei be given in a $D$-dimensional space with a
distance measure (in all of our applications, we use the isotropic
`Pythagoras' measure). Partition this space by assigning every one of
its points to its nearest nucleus. All the points in space assigned to
a particular nucleus $n$ form the {\it Voronoi region\/} of nucleus
$n$ (if $D=2$, the region is often called {\it Voronoi tile\/}). By
definition, the Voronoi region of $n$ consists of all points of the
given space that are at least as close to $n$ (according to the
distance measure of that space) as to any of the other nuclei. 
The set of all points that have more than one nearest nucleus constitute the {\it Voronoi
  diagram\/} of the set of nuclei \citep{edelsbrunner, okabe, orourke,
  vedeljensen}. For
clarity, most of our explanatory illustrations will use $D=2$ (see Fig.~\ref{fig:VoronoiDelaunay} for an
example in the plane), but our
applications have $D=3$.

The set of all points that have exactly two nearest nuclei $n_1,
n_2$ is the {\it Voronoi wall\/} between these nuclei. If $D=2$, such
a wall is a line; with $D=3$ it is a plane; and so on. In our
astrophysical applications, $D=3$, in which case (pathological
configurations excepted) triples of planar walls come together in {\it
  lines}, the points of which have three nearest nuclei; the
lines, finally, join in quadruples at {\it nodes}, single points that
have four nearest nuclei.
A Voronoi wall thus separates the Voronoi regions around two nuclei. The
connection between these nuclei, called an {\it edge}, is the
geometric dual of the wall. The set of all such edges is the {\it
  Delaunay triangulation\/}~\citep{Delone1934} of the point set. The Delaunay triangulation can be defined as the unique triangulation of nuclei for which the interior of a Delaunay simplex (triangle in two dimensions) contains no other nuclei.

In our algorithm, no other connections between nuclei are allowed, hence only adjacent Voronoi regions are connected. The expectation value for the number of these {\it Delaunay neighbours}, $\Lambda$, is 6 in 2D and 15.54$\cdots$ in 3D. 
Voronoi regions based on an
isotropic distance measure are convex; for a point process containing
$N$ nuclei, the number of geometric entities (walls, edges, nodes,
etc.) is ${\cal O}(N)$. 

Most important for our application to transport theory, is that the
Delaunay triangulation has a minimax property: this triangulation is
the one that has the largest smallest angle between adjacent triangle
edges. That is to say, of all possible triangulations of a given point
set, the Delaunay triangulation is the one that maximises the
expectation value of the smallest angle of its triangles.
Colloquially: the Delaunay triangulation has the least `sliver-like'
triangles, and the most `fat' triangles.

Another advantageous property of the Delaunay triangulation is that every edge pierces the wall between the Voronoi regions it connects at right angles. This justifies the use of Eq.(\ref{eq:RTtau}) where the spatial derivative is taken along the ray.

\subsection{Three types of transport}

In SimpleX, transport is always between neighbouring Voronoi cells (they are the ones connected by the Delaunay triangulation). On average every nucleus has 6 neighbours in 2D and 15.54$\cdots$ neighbours in 3D. When photons travel through a cell, the optical path length, $l$, is taken to be the average length of the Delaunay edges of that cell. If the number density of atoms in the cell is given by $n$ the fraction of photons that are removed from the bundle, $N_{\mathrm{rem}}$, is given by
\begin{equation}
N_{\mathrm{rem}} = N_{\mathrm{in}} e^{-n \sigma l}
\end{equation} 
where $\sigma$ is the total cross section which may be due to multiple extinction processes.

In general, extinction can be subdivided into absorption and scattering. Radiation that is removed by absorption processes will in general change the temperature of the medium and change its chemical state but it need not be transported further at this point.

Photons that are removed from the bundle due to scattering should propagate to neighbouring cells, either isotropically or with a certain directionality\footnote{As we will see in Sec.~\ref{sec:weighting_scheme_diff}, anisotropic scattering processes can be simulated straightforwardly by assigning weights to the outgoing edges so that more radiation is transported in the appropriate directions.}. This can be accomplished using the \emph{diffuse transport} method schematically depicted in the left panel of Fig.~\ref{fig:transport} and explained in more detail in Sec.~\ref{sec:diffuse}. 

The radiation that is not removed from the bundle, however, needs to travel straight on along the original incoming direction. For optically thick ($\Delta \tau > 1$) cells we simulate this using \emph{ballistic transport}  (see the centre panel of Fig.~\ref{fig:transport} and Sec.~\ref{sec:ballistic}). 

I n regions of the grid where the cells are optically thin ($\Delta \tau < 1$), ballistic transport becomes too diffusive and we need to resort to \emph{direction conserving transport} or DCT (see the right panel of Fig.~\ref{fig:transport} and  Sec.~\ref{sec:DCT}). 

We now proceed by describing these transport methods in more detail and show how they are combined in a general simulation.

\subsubsection{Diffuse transport}\label{sec:diffuse}

\begin{figure}[!ht]
   \begin{center}
     \includegraphics[width=0.45\textwidth]{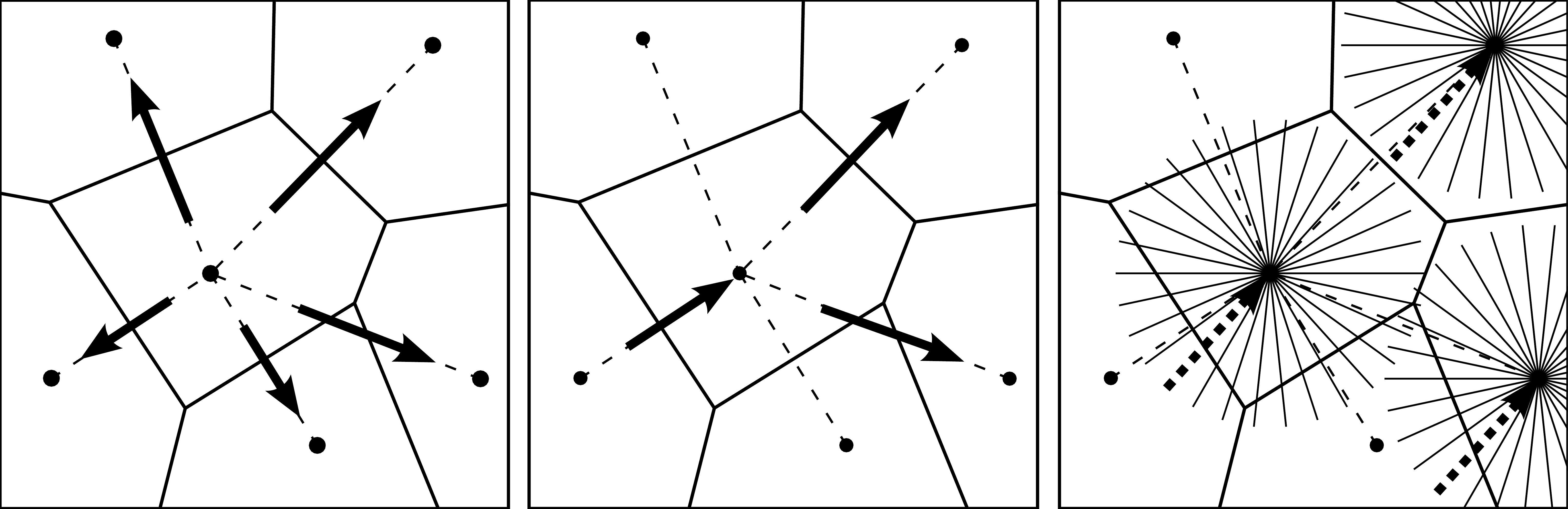}
     \caption{ Three principal means of transport supported in \emph{SimpleX\/}. Left panel: Diffuse transport,  photons from the incoming edge (not shown) are distributed outward along all edges (including the incoming edge). Centre panel: Ballistic transport, photons are transported along the D edges most straightforward with respect to the incoming direction. Right panel: Direction conserving transport,  photons (indicated with the dotted arrows) are transported as in ballistic transport but their direction is stored indefinitely in a global set of solid angles.
     }\label{fig:transport}
   \end{center}
 \end{figure}

We start with the description of conceptually the simplest form of transport implemented in \emph{SimpleX\/}: at each computational cycle, which we will call \emph{sweep}\footnote{Although in our current implementation each photon takes exactly one step per sweep, this need not be the case in general.} in the remainder of the text, the content of each nucleus is distributed equally among its neighbouring nuclei (see the left panel of Fig.~\ref{fig:transport}.)
We call this kind of transport \emph{diffuse} as it
has no memory of direction. In a homogeneous distribution of nuclei,
the transported quantity will diffuse outwards, spreading spherically
from the position of a source. This type of transport is appropriate for photons that are either scattered diffusely or absorbed and re-emitted in random directions.

\subsubsection{Ballistic transport}\label{sec:ballistic}

Now imagine a group of photons that is transported along a Delaunay
edge to a certain nucleus. Suppose that the nucleus represents a finite optical depth. A fraction of the photons will be removed from the group by the interaction and another fraction will fly straight onward. Diffusive transport is not suited
to describe this behaviour, so we introduce \emph{ballistic} transport.  

In the ballistic case, the incoming direction of the photons is used to
decide the outgoing direction (introducing a memory of one step into the past). In the generic Delaunay triangulation
there is no outgoing edge parallel to the incoming one, so the
outgoing photons are distributed over the $D$ most straightforward
directions, where $D$ is the dimension of the propagation space (see centre panel of Fig.~\ref{fig:transport}.)
As such, we ascertain that for an isotropically radiating source the complete `sky' is filled with radiation because the opening angle associated with each edge on average corresponds to $2\pi/\Lambda$ or $4\pi/\Lambda$ in two and three dimensions respectively\footnote{Another way to look at this is that, for an isotropically emitting source, the number of nuclei that receive radiation must scale as the square (cube) of the travelled distance for two (three) dimensions.}.

\begin{figure}[!ht]
  \begin{center}
    \includegraphics[width=0.45\textwidth]{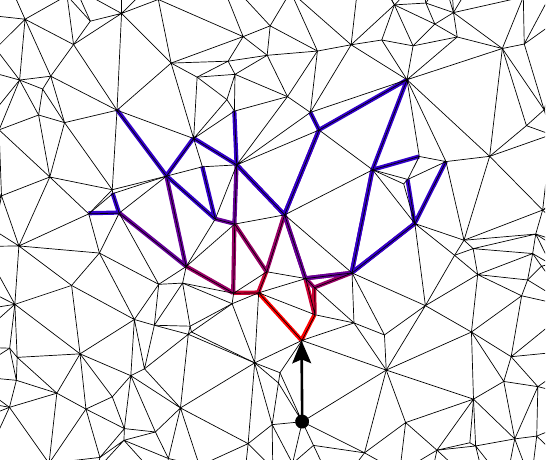}
    \caption{ Example of decollimation in the plane for five ballistic steps. The arrow indicates the initial influx of photons. According to the ballistic transport mechanism, photons are transported along the D most straightforward edges with respect to the incoming direction. The color coding is as follows: with every step the photons get a color that is less red and more blue. As the angle between adjacent edges is large in 2D (60 degrees on average), the bundle loses track of its original direction in only a few ballistic steps.}\label{fig:decollimation}
  \end{center}
\end{figure}

Note that due to the random nature of the directions in the Delaunay grid, radiation will tend to lose track of its original direction after several steps, a property we will call \emph{decollimation} as it will steadily increase the opening angle of a beam of radiation as it travels along the grid (see Fig.~\ref{fig:decollimation}. This property renders 
ballistic transport appropriate for highly to moderately optically thick cells only, where just a negligible amount of radiation has to be transported further than a few steps. If we take unity as a lower limit for the optical depth of a cell for which ballistic transport is used, at every intersection a fraction of $(1-1/e)$ of the photons gets absorbed and the
cumulative average deflection (decollimation) $\theta$ becomes
\begin{equation}
  \theta=
  \theta_D\sqrt{\sum_{n=1}^{\infty}\frac{1}{e^n}}=
 \theta_D\sqrt{\frac{1}{e-1}}\approx 0.76\,{\theta_D} \label{eq:deflectionLimit}
\end{equation}
where $\theta_D$ is the decollimation angle per ballistic step. In Sec.~\ref{sec:ball_transp_intr} we will measure  $\theta_D$ and describe the consequences of decollimation in more detail.

\begin{figure}[!ht]
  \begin{center}
    \includegraphics[width=0.45\textwidth]{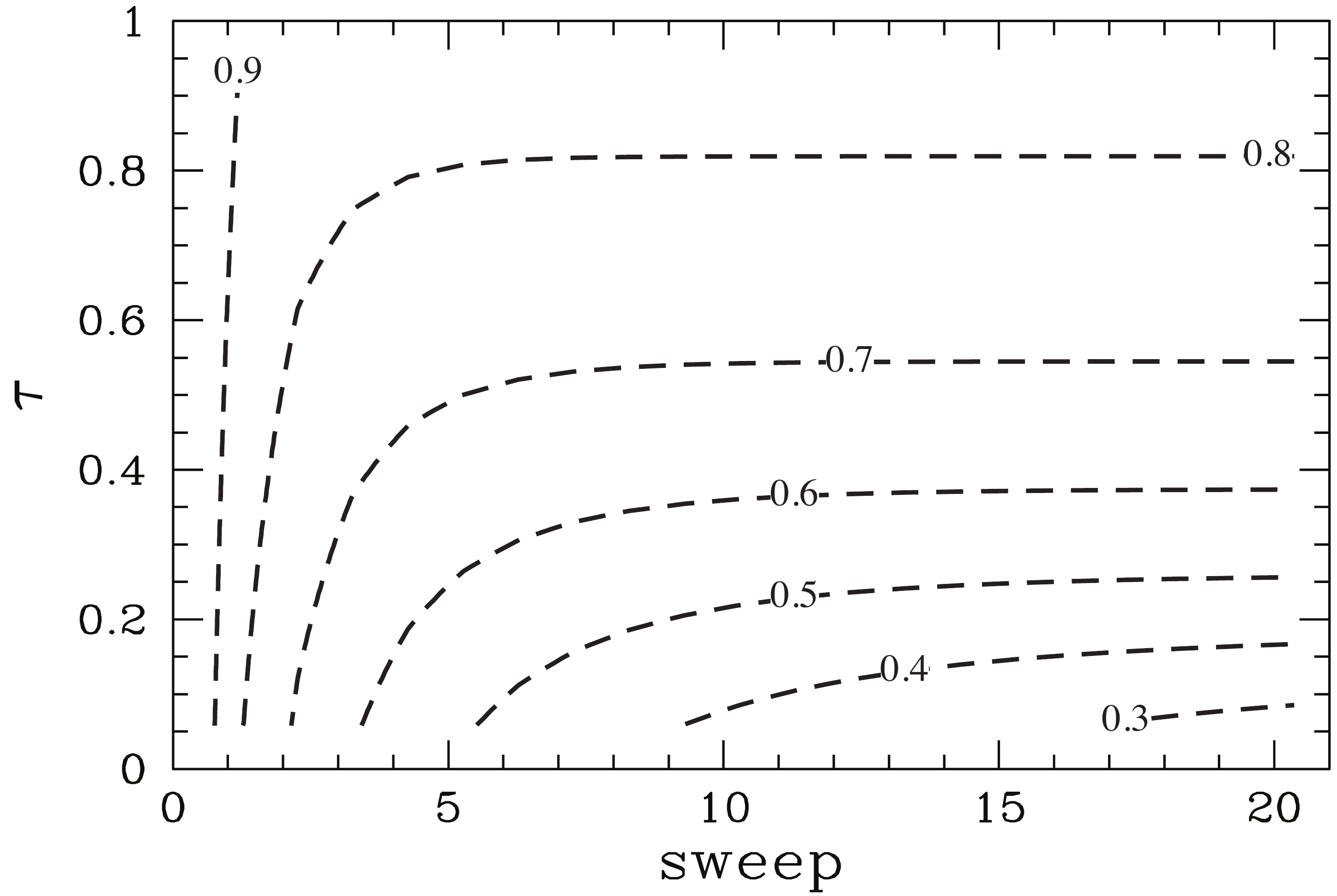}
    \caption{Fractions of photons (contours) with deflection angle within 45$^\circ$\,of the initial direction as a function of the number of ballistic steps (abscissa) and optical depth (ordinate).  From the figure it is evident that decollimation of photons is only potentially problematic when the cells are very optically thin. }\label{fig:fractions}
  \end{center}
\end{figure}

A visual extension of the statement given by Eq.(\ref{eq:deflectionLimit}) is shown in Fig.~\ref{fig:fractions}. From the figure it is evident that the fraction of photons that (ever) gets a deflection larger than 45$^\circ$ centred around the initial direction falls off sharply with the optical depth of a cell. Only for optically thin (say 0.2) cells the fraction of photons whose deflection stay under 45$^\circ$ is lower than 0.5.

In a realistic cosmological simulations, the effect of decollimation results in diffusion which softens shadows behind opaque objects (e.g. filaments and halos). The diffuse radiation field will penetrate into the opaque objects and ionise the high density gas inside. This results in ionisation of dense structures at too early times and the stalling of the ionisation front farther from the source. This clearly necessitates the introduction of a means of transporting photons in the optically thin regime.

\begin{figure}[!ht]
  \begin{center}
    \includegraphics[width=0.45\textwidth]{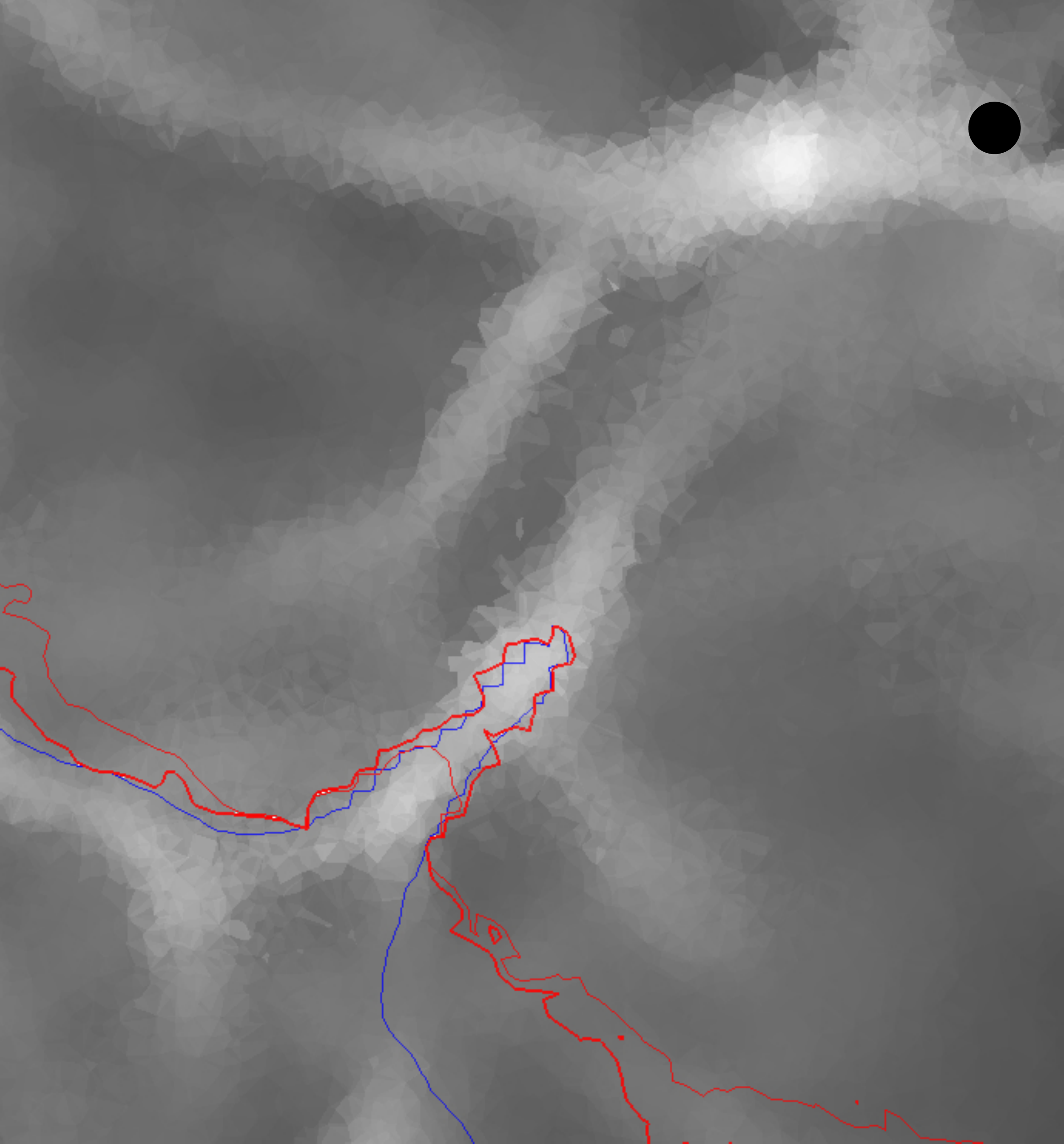}
    \caption{Shadow behind a dense filament irradiated by ionising radiation (indicated as a black dot). The hydrogen number density is plotted logarithmically in greyscale and ranges between $2.8\times10^{-5}$ to $0.12$ cm$^{-3}$. The red and blue lines indicates the contour where the neutral fraction of hydrogen is 0.3. The thin and thick red lines are for ballistic and direction conserving transport (DCT) respectively. The blue line is obtained with the C$^{2}$-ray code \citep{2006NewA...11..374M}. The contour for DCT shows a sharp 'shadow' (in good agreement with the C$^{2}$-ray result) where the dense filament is left neutral whereas ballistic transport results in a more diffuse, softer, shadow. Moreover, the ballistic ionisation front stalls with respect to the DCT result. This is another symptom of spurious diffusion: the positive radial component of the diffuse radiation is smaller than it should be resulting in more ionisations close to the source and less flux into the ionisation front. 
    }\label{fig:filament}
  \end{center}
\end{figure}

\subsubsection{Direction conserving transport}\label{sec:DCT}

If the cells are optically thin, the loss of directionality introduced by ballistic transport over many steps becomes prohibitive and we switch to \emph{direction conserving} transport (DCT for short). Here the radiation is confined in solid angles corresponding to global directions in space. So, if a photon has been emitted in a certain direction associated with a solid angle, it will remember this direction while travelling along the grid. 
This effectively decouples the directionality of the radiation field from the directions present in the grid (see right panel of Fig.~\ref{fig:transport}.) 

Note, however, that the transport of photons still happens along the three most straightforward Delaunay edges of the grid, where 'straightforward' is now with respect to the global directions of the solid angles\footnote{DCT can thus be viewed as ballistic transport with complete memory of direction.}. These straightforward edges may lead to nuclei that lie outside the solid angle associated with the direction of the photons. 

In this sense, we now have two types of angular resolution in our method. The first is related to the size of the solid angle in which radiation is confined spatially which is set by the number of Voronoi neighbours of a typical nucleus (15.54$\cdots$ in 3D).  We will call this {\it spatial resolution} and we stress that it depends solely on the nature of the Delaunay triangulation and as such is not adjustable. The second type of angular resolution is set by the global division of the sky in arbitrarily many directions (not necessarily a constant along the grid) and we will call this the {\it directional resolution}. 
For the tests presented here, we use 40 directions (a directional resolution of 40) implying a solid angle of $\pi/10$ sr for each unit vector. 

A technical consequence of the approach sketched above is the need to divide space into equal portions of solid angle whose normal vectors are isotropic. There are many ways to divide the unit sphere into equal patches but the requirement of isotropy cannot be met exactly in general. 

For maximal flexibility in angular resolution we have constructed sets of unit vectors that are distributed isotropically in angle using a simulated annealing scheme. 

One major drawback of the introduction of global directions in the simulation is that they will give rise to artifacts much like the ones observed in hydrodynamical simulations on regular grids. This in turn must be counteracted by randomisation of these global directions at appropriate time intervals which makes DCT computationally relatively expensive.
 
\subsection{Combined transport}

The three means of transport described above are applied simultaneously in general. If  the total optical depth $\tau_{\mathrm{tot}}$ for a given cell is due to multiple extinction processes
\begin{equation}
\tau_{\mathrm{tot}} = \sum_{i} \tau_{i},
\end{equation}
the fraction, $f_{i}$, of the incoming ray of photons that will be removed by process $i$ is given by
\begin{equation}
f_{i}\equiv \frac{\tau_{i}}{\tau_{\mathrm{tot}}}.
\end{equation}
Depending on the physical nature of this process, the photons are either removed from the bundle (e.g. to heat up the medium) or redistributed isotropically (or with some directionality) using diffuse transport. The remaining photons (those that are not removed from the ray) are transported with either ballistic or direction conserving transport, depending on the total optical depth of the cell (ballistic if $\tau_{\mathrm{tot}} \geq 1$ and DCT if $\tau_{\mathrm{tot}} < 1$). The fraction of the incoming photons that is treated with one of the three transport methods is thus fully determined by the grid.

\section{Anisotropy and its consequences}\label{sec:anis_its_cons}

The `ideal' properties of the Delaunay triangulation notwithstanding, the probability density distribution of the angle between adjacent Delaunay edges is quite
broad \citep[][Sec.5.5.4]{1987A&A...184...16I, okabe}, so that, even though the average
triangle is the `fattest' possible, many `thin' triangles will occur.
This situation may be changed by iteratively adapting the underlying
point process, in such a way that each nucleus comes to coincide with
the centre-of-mass of its Voronoi region \citep[see e.g. ][for such an algorithm]{Lloyd1982}, resulting in so-called Centroidal Voronoi Tessellations~\citep[see e.g.][]{du:637}. This procedure produces
`most spherical' Voronoi regions (actually hexagons if $D=2$), but is
in general far too costly for practical computations in which the
triangulation must be re-computed frequently. Moreover, the shifting
of the nuclei implies that the connection with the physical properties
of the underlying medium is no longer entirely faithful. Another reason to refrain from the use of Centroidal Voronoi Tessellations is the introduction of regularity and hence symmetry in the grid. In two dimensions, for instance, the hexagonal Voronoi cells tend to align, introducing globally preferential directions in the transport of radiation. This will inevitably give rise to artifacts in the radiation field transported on such a grid.

Although one cannot do better than using the Voronoi-Delaunay
construction on a random (Poisson process) point set, it is not
obvious that this still holds when the point set is inhomogeneous or
anisotropic. Inhomogeneity is, of course, the property we will
encounter in all practical cases. The probability distribution of the
directions of Delaunay edges on a Poisson nucleus is
isotropic, but this is no longer the case when the distribution of the
nuclei is structured.

This inherent anisotropy of the graph introduces a bias that is the cause
of some undesirable effects. It is these effects, and their treatment,
that we will consider here. We have tried in vain to locate a
mathematical treatment of related questions, such as: [a] Is the
Delaunay triangulation the one that maximises the isotropy of the
edges emanating from a given nucleus? [b] Can a process be found that
adjusts the positions of the nuclei in such a way as to increase the
isotropy of the edges? Concerning [a], we conjecture that the answer
is yes, because (at least superficially) that would seem to follow
from the minimax property of the angles between the edges. As to [b],
we have looked for a procedure analogous to the `centre-of-mass'
stratagem mentioned above, but have been unable to find one.

Thus, we must face the consequences of the anisotropy bias on
inhomogeneous point processes. We note in passing that such a bias
will exist locally even in the case of a Poisson process for putting
down the nuclei, because due to shot noise every instance of a random
point process is locally anisotropic, even if it is homogeneous and
isotropic in the mean.

\subsection{Error measure}\label{sec:quant-meas}

For a homogeneous Poisson distribution of nuclei, the expectation value for the number of Delaunay neighbours (and thus edges),  $\Lambda$,  is 6
in two- and $15.54\cdots$ in three-dimensional space. These Delaunay edges have no preferential orientation and their statistical properties are well known \citep[e.g.][]{okabe}.

Now consider the case that the distribution of nuclei is
inhomogeneous. Spatial gradients then appear in the density of nuclei,
$n(\vec{x})$, and the Delaunay edges connecting these nuclei are no
longer distributed evenly over all possible orientations. At a given
nucleus there will be, on average, more edges pointing towards
high-density regions than away from them.

This can be quantified as follows. Without loss of generality we may
assume a distribution of nuclei that has a gradient in some fixed
direction $x$ (provided that the characteristic length scale of the
gradient is much larger than the mean distance between nuclei, a
provision which we will assume to be fulfilled from now on). 

 Take a cross section perpendicular to the direction of the gradient
 through the box at arbitrary position $x_0$. The resulting plane with
 surface $S$ will be pierced by Delaunay edges connecting nuclei on
 either side of the plane.

The number of edges piercing the plane can be estimated as follows.
The local density of edges is the product of the number density of nuclei
multiplied with $\Lambda$. 
An edge is able to pierce the plane when two requirements have been fulfilled:
\begin{enumerate}
\item Its projected length must be larger than the distance between its originating nucleus and the plane.
\item It must point in the correct direction.
\end{enumerate}
The expectation value of the Delaunay edge length, ${\cal D}$,
 is given by
 \begin{equation}
   {\cal D}=\xi n(x)^{-1/3} \,,
   \label{eq:edge_length} 
 \end{equation}
 in three dimensions, where
 $\xi=1.237\cdots$~\citep[][Eq.(3.22)]{2007PhDT1R}.
 An edge emanating from a nucleus at position $x$ thus will only pierce
 the plane if its projected length exceeds $\Delta x\equiv |x-x_0|$. Hence, the first requirement is fulfilled when
 \begin{equation}
   \Delta x\leq{\cal D} \cos\theta\,,
   \label{eq:plane_dist} 
 \end{equation}
 where $\theta$ is the angle between the edge and the direction of the
 gradient. The orientation of these edges is random\footnote{The
   orientation might be correlated with the direction of the gradient
   but we will neglect this at the moment as it would only influence
   our results in second order.} so we must average the cosine over a
 half sphere which yields a factor of one half. 

 The effective length of the edges is thus ${\cal D}/2$ which can be interpreted as follows.
 Only nuclei inside a slab of thickness ${\cal D}$, centred at $x_{0}$
 contribute to the density of piercing edges. 
 
The second requirement effectively excludes half (up to first order) of the edges because they point away from the plane. This statement is equivalent to noting that every piercing edge connects exactly two nuclei at opposite sides of the plane. Including both factors we find that the number of piercing lines, $N_{\mathrm{p}}$, is given by
 \begin{equation}
   N_{\mathrm{p}}(x) = \frac{n(x)\Lambda{\cal D}S}{4}.
   \label{eq:total_piercing} 
 \end{equation}
 The surface density of lines piercing the slab, $\sigma(x)$, is now
 simply defined by
 \begin{equation}
   \sigma(x) = \frac{N_{\mathrm{p}}}{S} = \frac{n(x) \Lambda {\cal D} }{4}
   \label{eq:surf_density} 
 \end{equation}
 The sought after fractional excess, $E(x)$, of parallel (with respect to the gradient) over
 anti-parallel Delaunay edges is thus given by differencing $\sigma(x)$ over a
 sufficiently small\footnote{We emphasize the discrete nature of this
   derivative by noting that $\Delta n={dn\over dx}\,\Delta x={\cal
     D}{dn\over dx}=\xi n(x)^{-1/3}{dn\over dx}$ etc.} interval $\Delta x$ and
 division by $\Lambda n(x)$
 \begin{equation}
   E(x) \equiv \frac{1}{\Lambda n(x)}{\Delta{\sigma(x)}\over \Delta x}  = \frac{\xi}{4 n(x)^{4/3}}{\Delta n(x)\over \Delta x}\,.
   \label{eq:excess} 
 \end{equation}
where we have used Eq.(\ref{eq:edge_length}) to eliminate ${\cal D}$.
The excess of edges pointing towards higher density region
may have a signifcant effect on quantities transported along
these edges.  In the next three sections we will describe and quantify
these effects. We note in passing that in many practical applications (also see Sec.~\ref{sec:application}) the point density $n(x)$ of nuclei is taken to be proportional
to a power $\alpha$ of the mass density $\rho(x)$:
\begin{equation}
n(x)\propto\rho(x)^\alpha\,.
\label{eq:nuclei_mass_density} 
\end{equation}
In that case, Eq.(\ref{eq:excess}) becomes
\begin{equation}
E(x)={\xi \over 4n(x)^{1/3}}{\Delta \log n(x)\over \Delta x}={\alpha {\cal D}\over 4}{\Delta\log \rho (x)\over \Delta x}\,.
\label{eq:left_right_density} 
\end{equation}
If $\alpha=3$, the length ${\cal D}$ is proportional to the mean free
path of the photons \citep{2006PhRvE..74b6704R}. In many respects,
this is the `ideal' case, because is in fact
`transport-homogeneous' as experienced by the photon; every step is of equal optical depth. However,
Eq.(\ref{eq:left_right_density}) shows that such an `ideal' case has a
gradient asymmetry that is three times worse than is the case if
$n(x) \propto\rho(x)$.

\subsection{Effects on diffusive transport}

In Sec.~\ref{sec:diffuse} we stated that for diffusive transport in a homogeneous distribution of nuclei, the radiation propagates spherically away from a source. On a graph corresponding to an inhomogeneous distribution of nuclei, however, this is not the case. The spreading of the transported
quantity will not be spherical anymore due to two effects:
\begin{enumerate}
\item The Delaunay edges are shorter when the nuclei are spaced more
  closely.
\item The orientation of the Delaunay edges is no longer isotropic:
  more edges point towards the overdense regions.
\end{enumerate}

\subsubsection{Physical slow down}

The first effect reduces the transport velocity and can be interpreted
as a physical phenomenon: if we identify the length of a Delaunay edge
with the local mean free path of the transported quantity (e.g.
photons), the shorter edges simply express the fact that we have
entered a region of increased optical depth where it takes more mean
free path lengths to traverse a given physical distance. It has been
shown~\citep{2006PhRvE..74b6704R} that identifying the average
Delaunay edge length with the local mean free path of the relevant
processes is a natural choice in the construction of the triangulation
and the observed behaviour is therefore both expected and physical.

\subsubsection{Drift}\label{sec:effects_diff_transp}

\begin{figure}[!b]
  \begin{center}
    \includegraphics[width=0.45\textwidth]{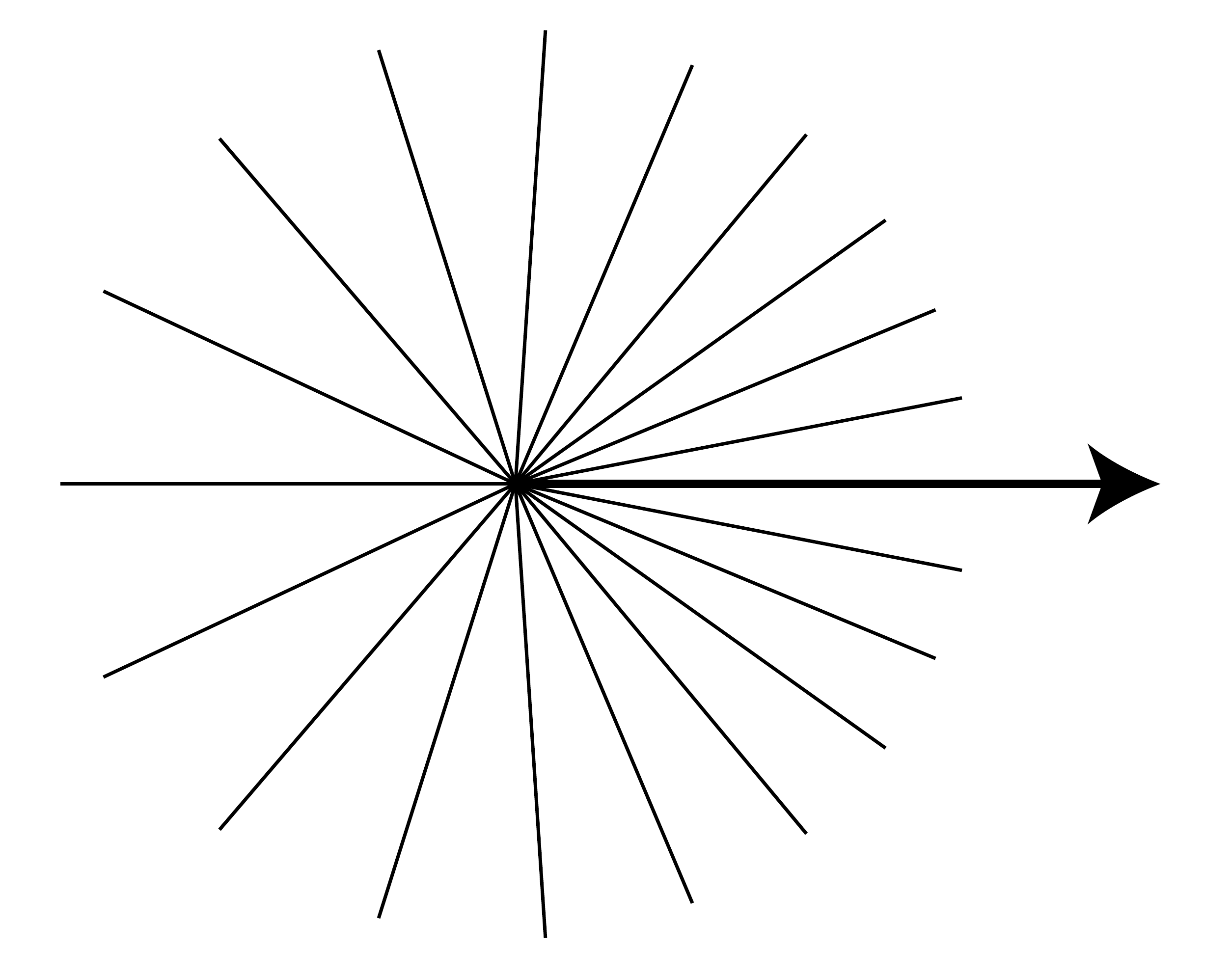}    \caption{Schematic example of a nucleus and its edges subject to a gradient in the number density of nuclei in the positive $x$-direction. More edges point toward the over-dens region (along the gradient). If every edge would transport an equal number of photons to neighbouring nuclei, the anisotropy of outgoing edges results in an unphysical net flow along the gradient (indicated by the arrow which is the vector sum of the edges scaled down by roughly a factor of three.)}\label{fig:diffusedrift}
  \end{center}
\end{figure}

The second effect, quantified by Eq.(\ref{eq:excess}), is an artifact
of the Delaunay triangulation itself and causes unphysical behaviour.
When there are too many edges pointing into the overdense regions, the
transported particles are deflected into those regions and the
direction of propagation tends to align with the gradient (see Fig.~\ref{fig:diffusedrift} for an example in the plane).  
We call this effect \emph{drift} and will now proceed to quantify its
consequences for diffusive transport.

A photon scattering at a nucleus has the following probabilities of
moving in the dense (subscript $d$) or underdense (subscript $u$)
direction:
\begin{eqnarray}
    p_d &  = & \frac{1}{2}+\frac{E(\vec x)}{2} \nonumber \\
    p_u &  = & \frac{1}{2}-\frac{E(\vec x)}{2} \,.
    \label{eq:under_over_prob}
\end{eqnarray}
The expectation value $d_D$ of the drift per scattering event is
therefore proportional to the $E(\vec x)$ part of an outgoing edge in
the direction of the dense region. The multiplication factor, which
can be found by integrating over a half-sphere, is $1/2$ since not all
edges point exactly to the right:
\begin{equation}
  d_D={{\cal D}\over 2} E(\vec x),
  \label{eq:drift}
\end{equation}
in which ${\cal D}$ is again the expectation value of the Delaunay edge
length (see Eq.(\ref{eq:edge_length})).

Say we set up a scattering experiment where a number of photons is
placed at one position of a triangulation with a density gradient. We
expect the photons to diffuse outward with an ever decreasing radial
velocity (the distance travelled,  $L$, scales with the root of the
number of steps) and at the same time drift towards the dense region
with a constant velocity.  There comes a time (or distance) at which
the drift is equal in magnitude to the diffusion radius. We define the
drift length, $L_{\mathrm{drift}}$, as the scale at which the diffusion distance
is equal to the distance travelled through drift. Roughly speaking,
diffusion dominates for $L<L_{\mathrm{drift}}$ and drift dominates for $L>L_{\mathrm{drift}}$.
Setting the diffusion distance equal to the distance travelled through
drift,
\begin{equation}
  {\cal D}\sqrt{N_E} = 
  {{\cal D}\over 2}E(\vec x)\, N_E,
  \label{eq:diffusion_drift}
\end{equation}
we find that the number of steps at equality is given by
\begin{equation}
N_{E}=4/E^{2}(\vec x). 
  \label{eq:Neq}
\end{equation}
Using Eq.(\ref{eq:excess}) and
Eq.(\ref{eq:edge_length}) gives an equality length of
\begin{equation}
  L_{\mathrm{drift}}(\vec{x}) ={8n(\vec{x})\over\mid\nabla{n(\vec{x})}\mid} \,,
  \label{eq:equality_length}
\end{equation}
independent of ${\cal D}$. Hence we have found a \emph{local}
expression for the relative importance of the unphysical drift.

For a given density distribution $n(x)$, the length $L_{\mathrm{drift}}(\vec{x})$ can
be evaluated everywhere\footnote{It may surprise the reader that
  values for $E(\vec x)$ and ${\cal D}$ are taken to be local while
  the argument seems to involve a domain in which these values could
  change. What actually happens is that we adopt the \emph{local}
  values for the \emph{whole} domain in
  Eq.(\ref{eq:diffusion_drift})}. This parameter can be interpreted as
follows. Say the minimum $L_{\mathrm{drift}}(\vec{x})$ for all $\vec{x}$ is $N$
times the box side length, then the drift can be at most $1/N$ of a
box side while the radiation scatters throughout the box. In any case
we must make sure that $L_{\mathrm{drift}}(\vec{x})$ is much bigger than the box
side length. 

Thus, even if the density contrast is
very small, it must not fluctuate too severely. Notice that it does
not help to increase the number of nuclei, since both $n(\vec{x})$ and
$\nabla n(\vec{x})$ in Eq.(\ref{eq:equality_length}) scale with that
number. 
Using more nuclei surely reduces the anisotropy per nucleus,
Eq.(\ref{eq:excess}), but because the number of steps to be taken increases accordingly, the effect simply adds up to the same macroscopic
behaviour. Only the shape of $n(\vec{x})$ is relevant.

In order to get a feel for the resulting constraint on the grid we consider the case for which $L_{\mathrm{drift}}(\vec{x})$ does not
depend on position. Demanding $L_{\mathrm{drift}}(\vec{x}) = {\rm constant}$ in
Eq.(\ref{eq:equality_length}) implies an exponential density
distribution. For such a distribution, the anisotropy is smeared out
maximally over the domain. In this case, if we want the drift to be
less than $1/8$ of the box length the density contrast must
be less than a factor $e \approx 2.71\cdots $ implying that the restrictions put by our
isotropy demands are rather stringent. 

\subsubsection{Clustering}\label{sec:effects_diff_transp_clust}

We have seen that the spurious drift for diffuse scattering puts
restrictions on the density contrasts that can be simulated by the plain implementation of \emph{SimpleX\/} introduced in Sec.(\ref{sec:effects_diff_transp}). In all of
this we have assumed that the scale of the density fluctuations is
comparable to the box size. This is not actually a restriction: if we are interested in a case
where the density fluctuations are of a much smaller scale than the
box size, we can just put an imaginary box around each density
fluctuation and use all the quantitative results from above. In doing so
we see that for any given `snapshot' too many photons will be present
in the local overdense regions, but if there are no overall density
contrasts on large scales, there will be no significant
macroscopic drift. The effect of the local drift on small scales can
still influence results on a large scale, however.
Not only does the drift influence the average position of the photons,
it also influences the standard deviation around this average.
Isotropic scattering maximises the spreading of photons, but in an
extreme case where for example at every nucleus 90\% of the photons move in the
same general direction, these will stick together for a longer period
and therefore the size of a `light cloud' will grow more slowly. This
is an effect that shows up if we have a highly fluctuating density
field on small scales. A case in point is the simulation of the
filaments of large scale cosmic structures.

Consider a density distribution that is homogeneous in the
{\it y}- and {\it z}-direction but highly fluctuating in the {\it x}-direction. For the
probabilities to travel into the dense or underdense regions we again
use Eq.(\ref{eq:under_over_prob}) but with the difference that now $d$ and $u$ no
longer denote global directions. We describe the
transportation process by a binomial distribution. If there are no
large scale density contrasts the drift will be zero on average, but
for the standard deviation we find:
\begin{eqnarray}
  \sigma(N) & = & \langle \sqrt{N\, p\, (1-p)} \, \rangle \nonumber \\
  & = &  \left \langle \sqrt{N\left(\frac{1}{2}+\frac{E(x)}{2}\right)\left(\frac{1}{2}-\frac{E(x)}{2}\right)} \, \right \rangle \nonumber \\
  & = &  {\sqrt{N}\over 2} \left\langle \sqrt{1-E^2(x)}\ \right\rangle \,
  \label{eq:clustering}
\end{eqnarray}
where $N$ is the number of sweeps. For small values of $E$, this
reduces to
\begin{equation}
\sigma(N) \approx {\sqrt{N}\over 2}\left(1-{\xi^2\over 32} \left\langle\left|
{n'(x) n(x)^{-4/3}} \right|^{2} \right \rangle\right) \,. 
  \label{eq:sqrtfactor_approx}
\end{equation}
This factor is always smaller than unity, indicating a reduction of the
spreading of the photons. This effect can be significant for
small-scale fluctuations with either a very high amplitude or a very
short length. The effect slowly becomes smaller if the
number of nuclei is increased. The value of $E(x)$ comes close to unity
or exceeds unity only if the characteristic density fluctuation length is smaller than a Delaunay length, which we had pointedly excluded.

\subsection{Effects on ballistic transport}\label{sec:effects_ball_transp}

For ballistic transport, similar problems arise as for diffusive
transport but the fact that the transport is locally anisotropic
complicates the picture. We first describe some properties of this kind of
transport on a homogeneous grid in order to appreciate
deviations from the 'normal' case later on.

We identify two distinct phenomena that photons travelling
ballistically may experience: deflection and decollimation. Deflection
is here understood as 'loss of direction', where the direction is given
by the vector sum of the three most straightforward directions.
Decollimation is defined to be the effect of the increase in opening angle
of a beam of photons as they are transported ballistically. 
The first phenomenon depends on the evolution of the vector sum of the three
most straightforward directions whereas the second phenomenon is
related to the angular separation between these directions
individually. In the case of a homogenous distribution of nuclei, the net effect of deflection will vanish because there is no preferential direction in the grid.

\subsubsection{Decollimation}\label{sec:ball_transp_intr}

As stated in sec.~\ref{sec:ballistic}, the angular resolution, and therefore the minimal opening angle, is a property of the Voronoi cell and is thus fixed for the chosen triangulation. To exemplify this we consider a typical Voronoi nucleus in 3D, connected to $\Lambda$ neighbouring nuclei. The expectation value for the solid angle, $\Omega$, subtended by each edge is thus
\begin{equation}
\Omega = \frac{4 \pi}{\Lambda}\approx \frac{\pi}{4}.
\label{eq:sol_ang}
\end{equation}

\begin{figure}[!b]
  \begin{center}
    \includegraphics[width=0.45\textwidth]{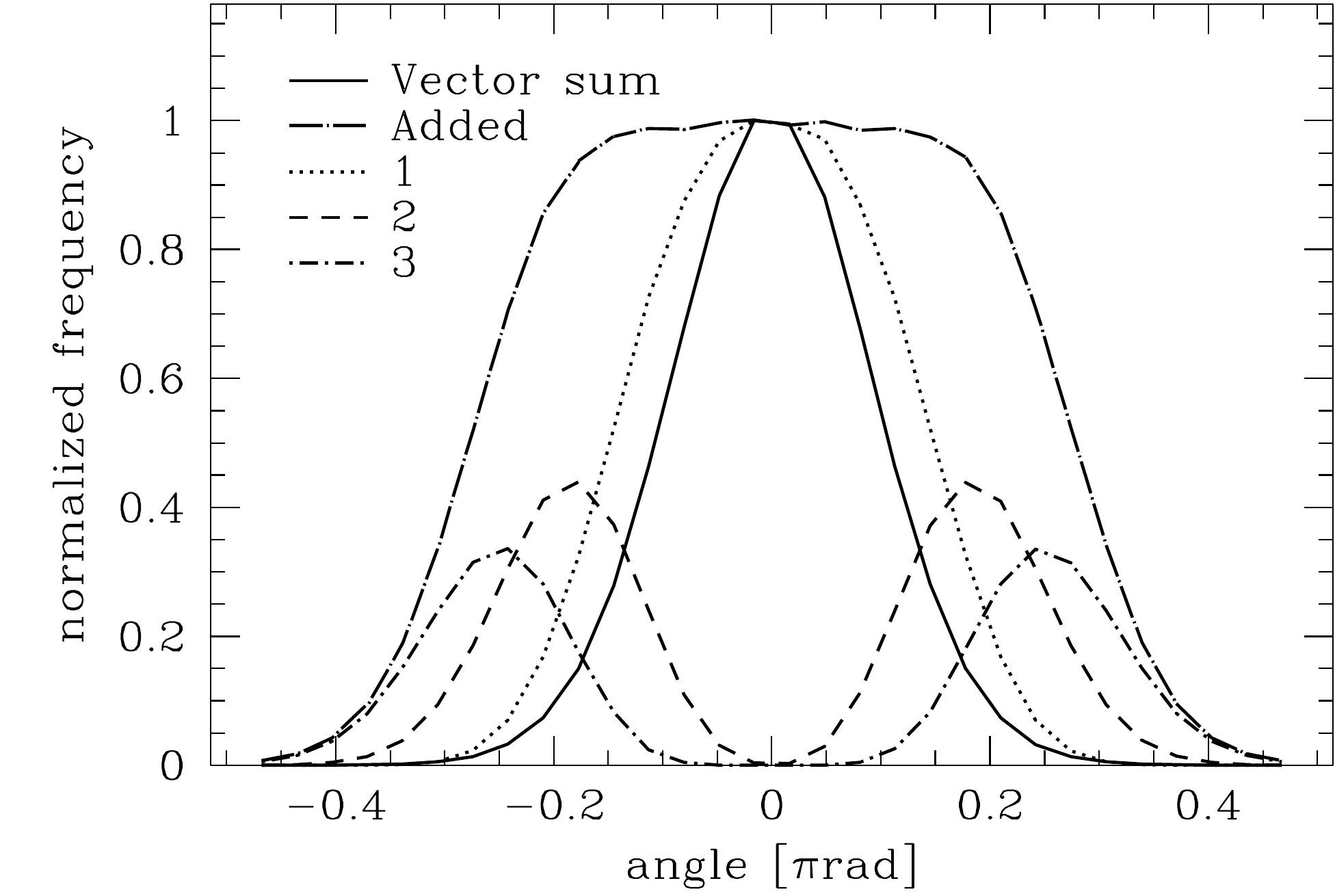}    \caption{Normalised distribution of angles between the incoming direction and the vector sum of the (three) most straightforward pointing edges
      (solid line) and the separate most straightforward edges (dotted, dashed and dot-dashed 
      lines) for a homogeneous distribution of nuclei. The distribution of the most straightforward edges added is shown as the long dot-dashed line with label `Added' and is related to the decollimation effect. The standard deviation (FWHM$/\sqrt{\ln 256}$) is about
     \bvalang for the added edges and \thetadef for the
      vector sum of the edges.}\label{fig:balldev}
  \end{center}
\end{figure}

In Fig.~\ref{fig:balldev} the distribution of angles is given for a
grid of 10$^{5}$ homogeneously placed nuclei in three dimensions.  For the
vector sum the average departure from the incoming direction is about
\thetadef (solid line). This implies that a photon loses all knowledge of its
original direcion in typically  (90/15)$^{2}= 36$ steps.

For the distribution of the three separate most straightforward edges
we find a standard deviation of about \bvalang which means that after $(90/39)^{2}\approx 5$  steps the photon has lost all memory of its original direction. Note that the width of the vector sum of the weighted edges is smaller than that of the most straightforward edge (indicated by `1'  in the figure) alone.

\begin{figure}[!ht]
  \begin{center}
    \includegraphics[width=0.45\textwidth]{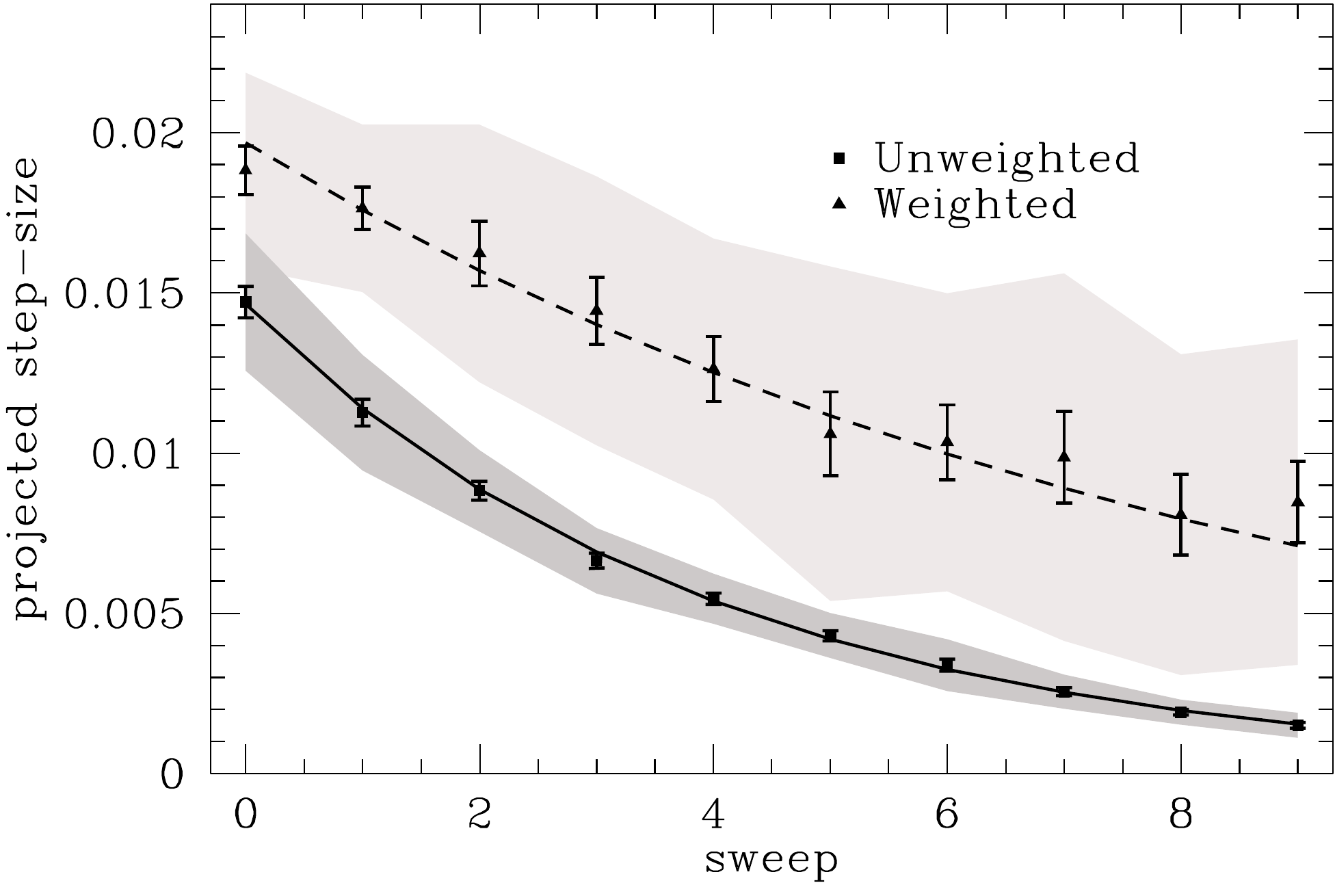}
    \caption{The projected step-size in the x-direction as a function of 
      sweeps together with a least squares fit for the unweighted ballistic transport (solid line) and with ballistic weights as described in Sec.~\ref{sec:weighting_scheme_ball} (dashed line). A region indicating unit standard deviation is shaded in grey around the markers. Error bars corresponding to $\sigma/\sqrt{N-1}$ where $\sigma$ the standard deviation and $N$ the number of experiments executed. For the uncorrected case the fitted value for the deflection angle $\theta_{D}$ in Eq.~(\ref{eq:fitfunc}) is \bvalang$ \pm $ \bvalerrproc. For the weighted case we find $\theta_{D}=$ \bvalangweights $\pm $ \bvalerrprocweights.
    }\label{fig:balldiff}
  \end{center}
\end{figure}

In general, the gradual loss of direction is not a major concern since most
sources emit isotropically anyway, but the effect becomes important
when many edges are traversed, because then the photons behave diffusively,
 producing an intensity profile that deviates from the correct $r^{-(d-1)}$-form.  Simulations in which
the mean free path for scattering or absorption is much smaller than
5 edges are fine in this respect, because then the ballistic photons
never enter this random walk regime (see also the discussion in
Sec.~\ref{sec:discussion}).
To observe the implications of this statistical statement in the
transport of photons we construct a triangulation of a random point
distribution containing 10$^{5}$ homogeneously placed nuclei in a cube of
unit dimension. In the centre we define a small spherical volume with
radius 0.05 and place a number of photons in each nucleus contained in
this volume. The photons are assigned a direction by sending them
along the edge whose direction is maximally parallel to one of the
coordinate axes (we will choose the positive x-axis). So, when
applying our ballistic transport method, the photons will start to
move in the positive x-direction with a step-size equal (on average)
to $\mathcal{D}\cos\theta_{D}$ where $\theta_{D}$ is the mean decollimation angle. Every step means a multiplication by a
factor $\cos\theta_{D}$, until after infinitely many steps the
effective step size in the x-direction is zero. In
Fig.~\ref{fig:balldiff} the step-size as a function of sweeps is shown
together with a fitting function of the form 
\begin{equation}
f(N_s)=A\cos(\theta_{D})^{N_s}\label{eq:fitfunc}
\end{equation}
where $N_s$ the number of sweeps. We find that in the case without ballistic weights (filled squares; see Sec.~\ref{sec:weighting_scheme_ball} for the description of ballistic weights) $\theta_{D}=\bval \, \pm \, $\bvalerrproc \, (\bvalang $ \pm$  \bvalerrproc) in accordance with the value obtained from the grid statistics
presented in Fig.~\ref{fig:balldev}.

\subsubsection{Deflection}\label{sec:effects_due_anis}

We are now ready to quantify the effect of anisotropy of the
triangulation on the ballistic transport of photons, which travel
along the three edges closest (in angular sense) to the incoming
direction. For the sake of simplicity we will estimate the deflection for radiation travelling along only one most straightforward edge and discuss the applicability to three edges afterwards. 

\begin{figure}[!ht]
  \begin{center}
    \includegraphics[width=0.2\textwidth]{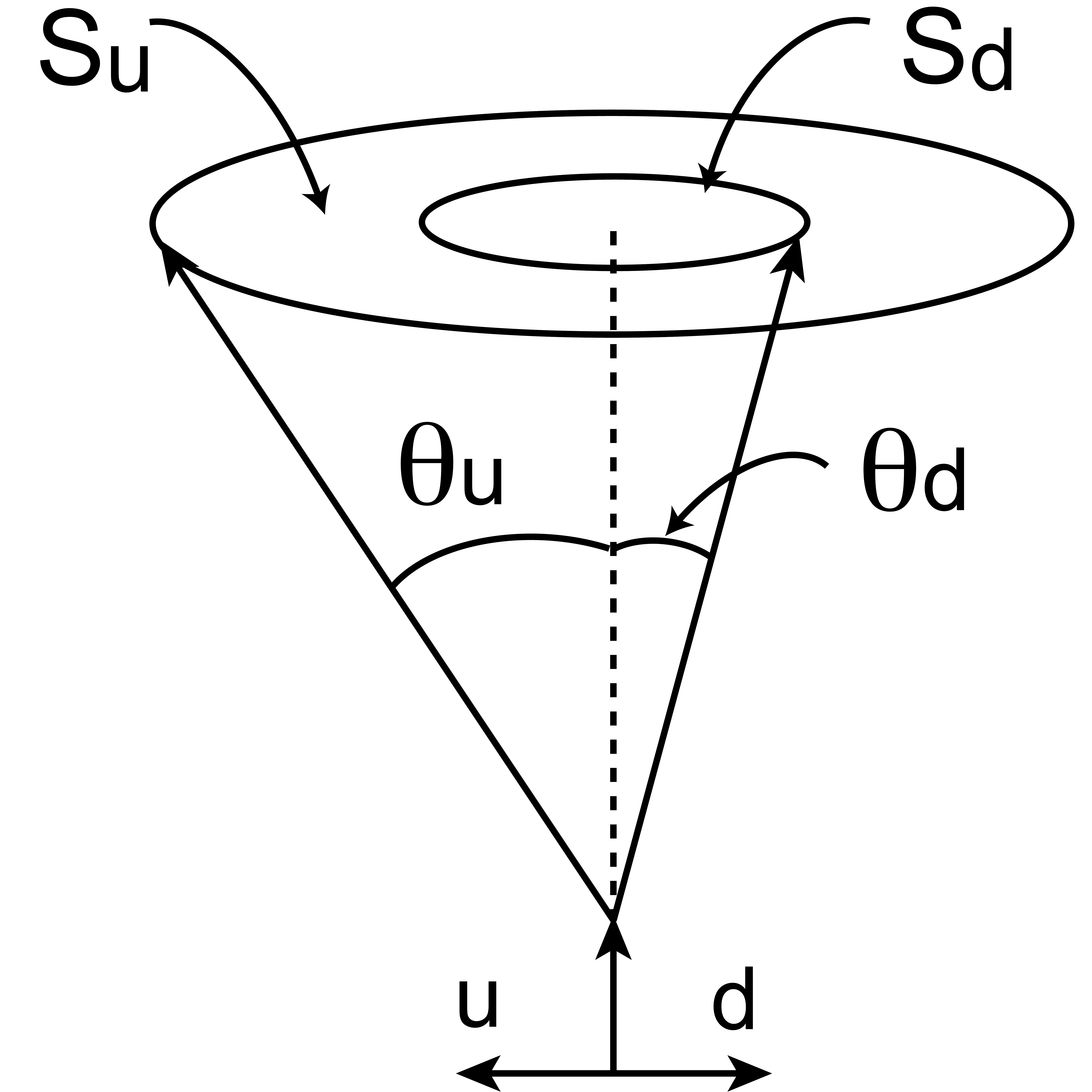}
    \caption{Geometry of radiation travelling perpendicular to the gradient direction. The deflection angle toward the under-dense region, $\theta_{\mathrm{u}}$, scales with the radius of the area S$_{\mathrm{u}}$ and similarly for the deflection angle towards the over-dense region, $\theta_{\mathrm{d}}$. 
    }\label{fig:deflection_geom}
  \end{center}
\end{figure}

Consider photons streaming perpendicular to the gradient direction (see Fig.~\ref{fig:deflection_geom} for the geometry of this situation.) The standard deviation of the deflection of the outgoing edge with respect to the incoming direction is typically \thetadef (see Fig.~\ref{fig:balldev}) for a homogeneous grid but will in general depend on the local value for the gradient. 
Referring to Eq.(\ref{eq:sol_ang}), the expected opening angle depends on the effective number of outgoing edges in that direction. In other words, as the anisotropy increases the number of edges pointing toward the over-dense region, the angular resolution increases accordingly. This motivates one to define the direction dependent number of outgoing edges, $\Lambda_{\mathrm{eff}}$, as
\begin{equation}
\Lambda_{\mathrm{eff}}(\phi) = \Lambda [ 1 + E(x) \cos \phi ],
\label{eq:sol_ang_eff}
\end{equation}
where $ \phi $ is the angle with the gradient direction. If we interpret the solid angle as a projected circular area on the unit sphere we can estimate the maximal deflection angle, $\theta_{\mathrm{d}}$, for ballistic transport as
\begin{equation}
\theta_{\mathrm{d}} = \arcsin \sqrt{
	\frac{ 4 }{\Lambda_{\mathrm{eff}} (\phi) }
	}.\label{eq:theta_def}
\end{equation} 
For a homogeneous distribution of nuclei, this angle evaluates to approximately $30^{\circ}$  with a typical value of \thetadef as expected from the analysis shown in Fig.~\ref{fig:balldev}. If we include this empirical factor of one half in Eq.(\ref{eq:theta_def}) and approximate the arcsine by its argument (correct within 1\% for angles smaller than $\pi/12$) we obtain
\begin{equation}
\theta_{\mathrm{d}}  = (\Lambda [ 1 + E(x) \cos \phi ])^{-1/2},
\end{equation}
where we have used Eq.(\ref{eq:sol_ang_eff}) to substitute for $\Lambda_{\mathrm{eff}}$. This result can be approximated to first order by 
\begin{equation}
\theta_{\mathrm{d}}  \simeq \Lambda^{-1/2} [ 1 -\frac{ E(x) \cos \phi}{2} ].
\end{equation}
To obtain the deflection per step over the grid, we average $\theta_{\mathrm{d}}$ over azimuthal angle while projecting along the direction of the gradient
\begin{eqnarray}
\theta_{\mathrm{eff}}  & = & \frac{1}{2 \pi}\int_{0}^{2 \pi} \theta_{\mathrm{d}} \cos \phi d\phi \\
\ & = & -\frac{E(x)}{4 \sqrt{\Lambda}}.\label{eq:theta_eff}
\end{eqnarray}
The sign of $\theta_{\mathrm{eff}}$ is negative, which means that the resulting deflection is into the region of lesser density. This result might surprise the reader at first. One could naively predict the photons to deflect into the direction of higher density (similar to the diffuse drift) but the situation is exactly reversed for ballistic deflection. 

Note that this effective deflection angle is an upper limit for the deflection encountered in a simulation because of the following. The effect is maximal for radiation travelling perpendicular to the direction of the gradient and this is the situation we have used as a starting point for the above derivation of $\theta_{\mathrm{eff}}$.  Secondly, the fact that $\theta_{\mathrm{d}}<\theta_{\mathrm{u}}$ will favour the selection of edges in the over-dens region effectively diminishing the deflection, an effect that we have neglected in the derivation above. To include this effect one would have to know the distribution of outgoing edges as a function of angle with the gradient, and assign probability to the selection of an edge accordingly. As we shall see in Sec.~\ref{sec:deflection}, the omission of this effect does not seem to be of much importance for our predictions.

Sending radiation along three edges rather than one will further decrease the effect of deflection. This is expected as the deflection of one edge is larger than that of the vector sum of three outgoing edges as we saw in the case of a homogeneous grid. 

The growth of the deflection when traversing the grid can be found by differencing Eq.(\ref{eq:theta_eff}) by the typical step-size, $\mathcal{D}$
\begin{equation}
\frac{\Delta \theta_{\mathrm{eff}} }{\Delta l} = \frac{1}{16 \sqrt{\Lambda}} {\nabla n(x)\over n(x)}.
\end{equation}

Analogous to the drift length of Eq.(\ref{eq:equality_length}) we can define a deflection length,  $L_{\mathrm{def}} $, at which the cumulative deflection is equal to say,  $\pi/4$, 
\begin{equation}
L_{\mathrm{def}} = \frac{4 \pi \sqrt{\Lambda} n(x)}{\nabla n(x)} \approx 50  \frac{n(x)}{\nabla n(x)}.\label{eq:L_def}
\end{equation}
Comparing with Eq.(\ref{eq:equality_length}) we see that both the
diffuse drift and the ballistic deflection give approximately the same
restrictions on the box size, the only difference being a factor 50
instead of 8. Again the interpretation is simple. If
$L_{\mathrm{def}}(\vec{x})$ is strictly larger than say five times the box
size, the deflection will be less than one-fifth of $\pi/4$ in
travelling across the box in a `straight' line. Remember that this
effect is added to the expected decollimation from
Sec.\ref{sec:ball_transp_intr}.

\section{Weighting schemes}\label{sec:weighting_scheme}

In this section we describe several possibilities to correct for the unphysical effects described in Sec.~\ref{sec:anis_its_cons}.

\subsection{Diffuse transport}\label{sec:weighting_scheme_diff}

\begin{figure}[!ht]
  \begin{center}
    \includegraphics[width=0.40\textwidth]{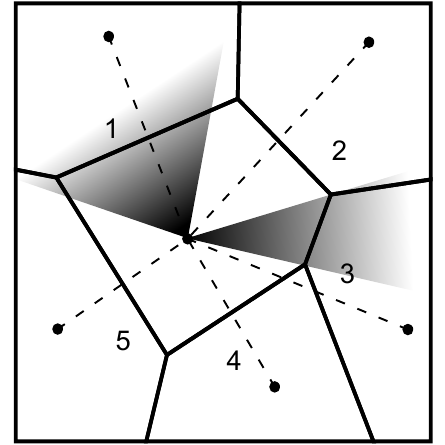}
    \caption{Solid angles for two edges in the case of the Voronoi weighting
      scheme (cell 3) and the icosahedron weighting scheme (cell 1).  In the case of the Voronoi scheme, the weight assigned to a Delaunay edge corresponds directly to the
      magnitude of the solid angle of its corresponding Voronoi cell.
      In the icosahedron scheme, the weight is proportional to the
      solid angle a Delaunay line occupies considering the angular
      vicinity of its neighbouring edges (in 2D the solid angle is thus bound by the two bisectors shown as dotted lines in the figure.)}\label{fig:solidAngle}
  \end{center}
\end{figure}

A straightforward cure for the problems addressed in 
Sec.~\ref{sec:effects_diff_transp} and Sec.~\ref{sec:effects_diff_transp_clust} (diffuse drift and clustering respectively) is to assign weights $w_i$ to the
edges emanating from a given nucleus in such a way that the anisotropy
vanishes. This means that the fractions of the quantity transported to
the neighbours are no longer equal to 1/$N$ but directly proportional
to the solid angle that the corresponding Voronoi face spans. We refer
to Fig.~\ref{fig:solidAngle} for an example in the plane, the
three-dimensional case is analogous.
We have explored this possibility and implemented three different
weighting schemes in our method, each with its specific virtues and
drawbacks:

\begin{enumerate}
\item Voronoi weights: based on the natural properties of the
  triangulation.  Advantages: automatically and `naturally'
  adapted to the physics of the transport problem. Disadvantages:
  inherently noisy.
\item Icosahedron weights: based on a division of the unit sphere
  using the icosahedron. Advantages: flexible (does not depend on type
  of triangulation), refinable. Disadvantages: computationally
  expensive.
\item Distance weights: based on the distance to a neighbour squared. Advantages: very fast. Disadvantages: Only empirical evidence of its correctness.
\end{enumerate}

\subsubsection{Centre of gravity weighting}

\begin{figure}[!ht]
  \begin{center}
  \includegraphics[width=0.3\textwidth]{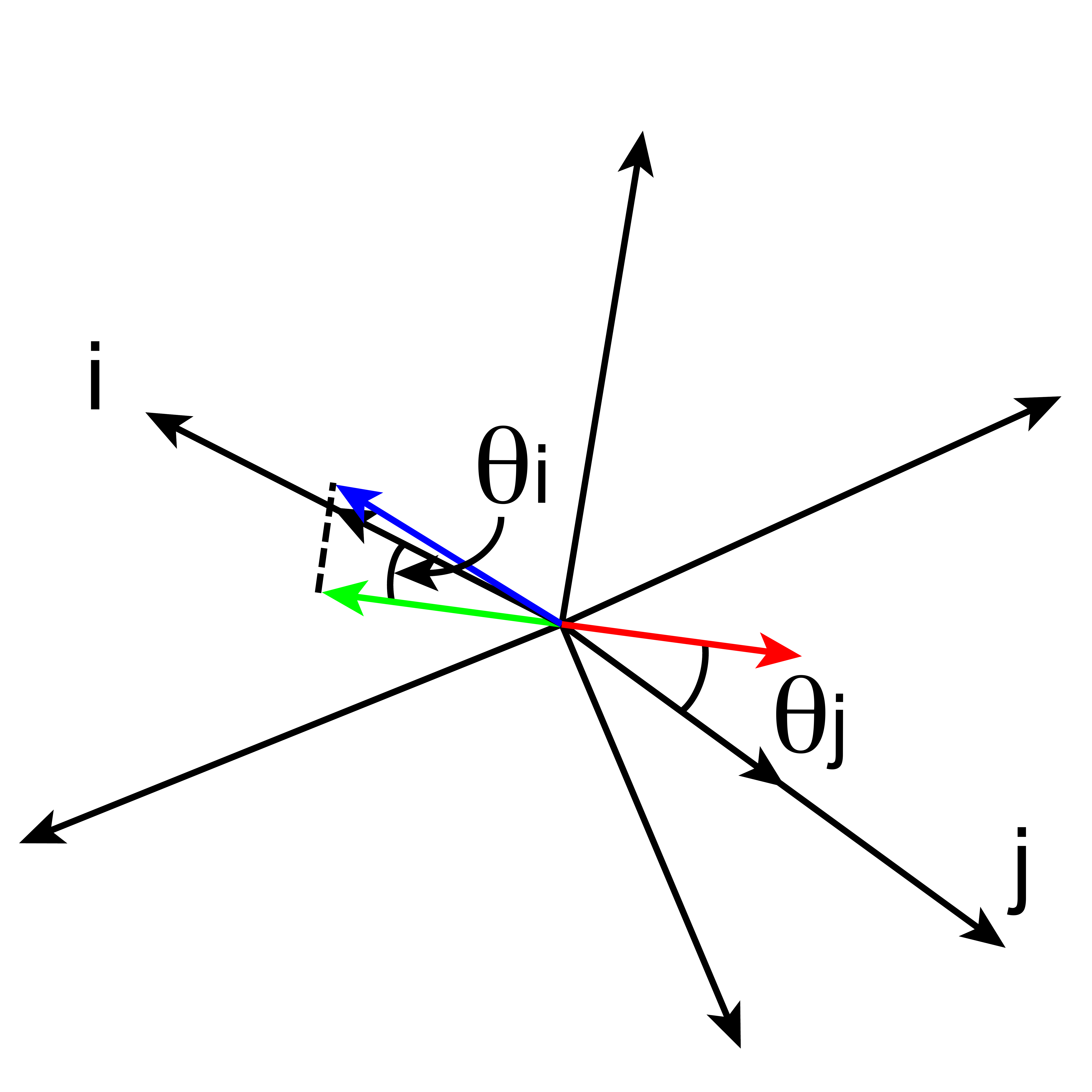}
  \caption{Geometry of the COG weighting procedure. In general, the vector sum of all (weighted) Delaunay edges (black arrows) is non-zero (red arrow). The Delaunay edges $j$ and $i$ are most parallel and anti-parallel to this vector sum respectively. Their weights are adjusted such that the vector sum vanishes in its current direction. To this end, the vector sum and its reflected counterpart (green arrow) are projected on $j$ and $i$, indicated with the smaller black arrows. Half of this projection is added to edge $i$ and half is subtracted from $j$. The resulting vector is shown in blue and it exactly cancels the vector sum. The adjustment of weights can still result in a non-zero (although smaller) vector sum in some other direction. The above procedure is repeated iteratively until the norm of the vector sum is smaller than a predefined tolerance.}\label{fig:COG}
    \end{center}
\end{figure}

In all three weighting methods, the weights sum up to unity, guaranteeing conservation
of photons. In addition the \emph{vector} sum
 of the weighted Delaunay
edges emanating from a nucleus should be zero to conserve momentum of
the radiation field. We obtain this by adjusting the weights of the
(two) edges most parallel and anti-parallel to the `centre of
gravity', $\Psi$, of the nucleus defined by
\begin{equation}
  \Psi = 
  \sum_i w_i \hat r_i
  \label{eq:mass_centre}
\end{equation}
such that in that direction the magnitude of $\Psi$ vanishes (see Fig.~\ref{fig:COG}.) Here $\hat r_i$ denotes the direction of the $i$-th Delaunay edge. We
repeat this procedure until the absolute value of $\Psi$ is smaller
than some given tolerance. In our experience, this procedure reaches a
(relative) tolerance of 10$^{-7}$ within twenty iterations for the
Voronoi method and about ten for the Icosahedron and Distance weighting schemes.

\subsubsection{Voronoi weighting}

The Voronoi weighting scheme uses the faces of the Voronoi cells to
calculate the $w_i$ for each nucleus by giving an estimate for the
solid angles subtended by the walls of the Voronoi regions, normalised
to unity. The weights are then
\begin{equation}
  w_i = {S_i\,d_i^{-2}\over\sum_j S_j\,d_j^{-2}},
  \label{eq:VorWeights}
\end{equation}
where $S_i$ denotes the surface of the Voronoi face perpendicular to
the Delaunay edge connecting the current nucleus and its $i$-th
neighbour and $d_i$ is the length of that edge (because the Voronoi
face cuts the Delaunay edge in the middle, $0.5 d_i$ is the distance
from the nucleus to the face).  Note that Eq.(\ref{eq:VorWeights}) is
an approximate expression neglecting projection effects for large
angles. With some computational effort, this estimate could be
refined. The solid angle subtended by a Voronoi face depends on the
size of the cell but also on the precise position of the nuclei.
Referring to Fig.~\ref{fig:solidAngle}, we see that cell `$3$'
has a relatively small surface because its nucleus is close to that of
neighbour `$4$' \emph{and} its nucleus is a bit further from the
central nucleus than that of neighbour `$4$'. If that last statement
were reversed, neighbour `$4$' would have the smaller surface and would thus get the smaller weight. This property may seem harmful at first but it is of stochastic nature,
meaning that some noise will be introduced but no systematic error. 

\begin{figure}[!ht]
  \includegraphics[width=0.45\textwidth]{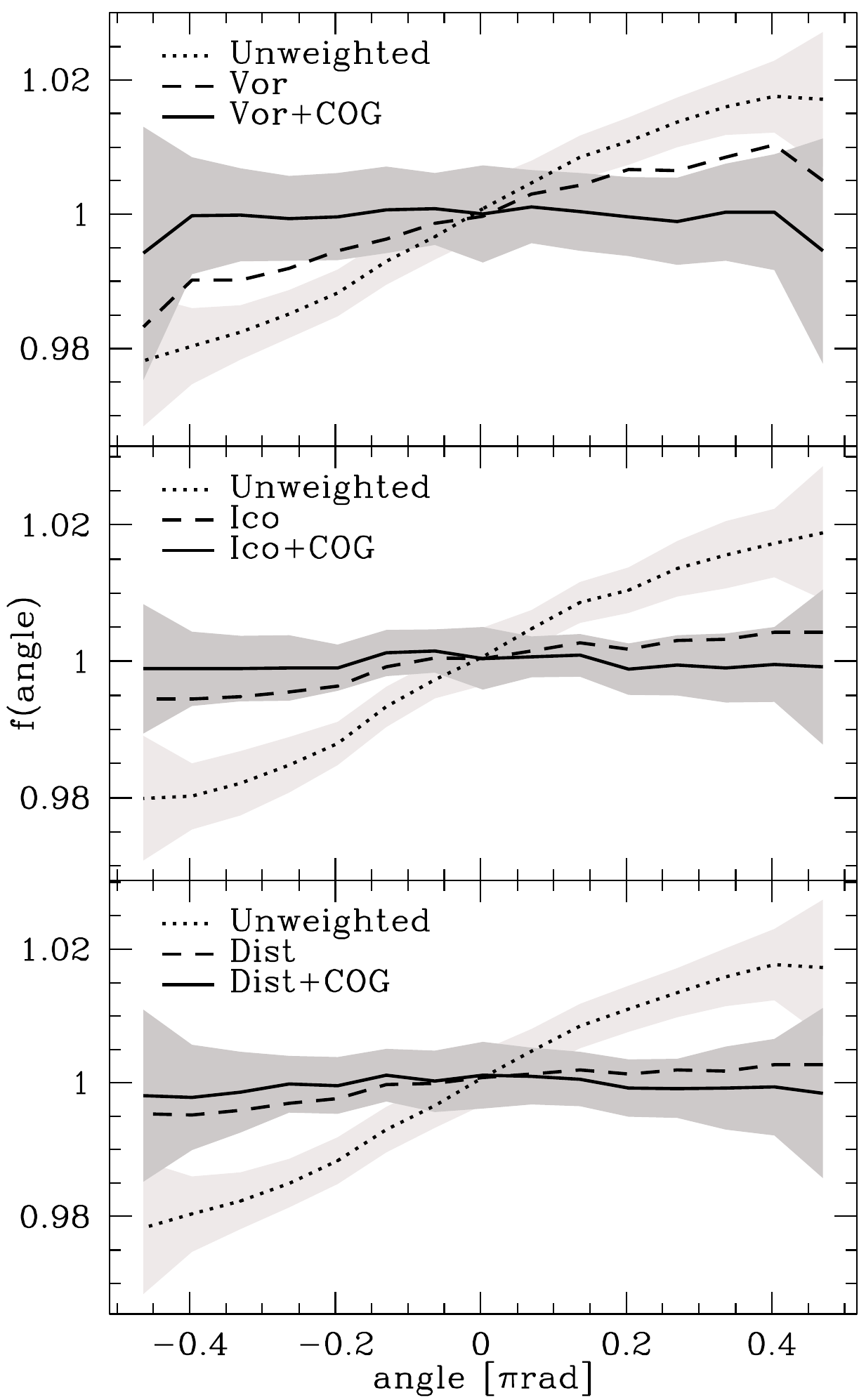}
  \caption{Fraction of edges, $f(\theta)$, with angle $\pi-\theta$ 
    with respect to the direction of the linear gradient as a
    function of $\theta$ for different cumulatively applied
    corrections with the Voronoi (top panel), Icosahedron (middle panel) and Distance weighting scheme (bottom panel). The results are
    averaged over 100 different realisations of the grid to suppress
    shot noise. Unit standard deviation regions around the 'Uncorrected' curve and the final result are shaded in light and dark grey respectively. As a result of the anisotropy of the triangulation
    more edges point towards the overdense region (to the right) in
    the uncorrected case. Note that due to normalization, the
    results at the extreme ends are subject to
    noise.}\label{fig:weights}
\end{figure}

To demonstrate our method we have constructed a Delaunay triangulation
of 10$^{5}$ nuclei in three-space with a linear gradient in $n(\vec{x})$
along the horizontal direction which runs from 0.005 on the left to
0.995 on the right. As a measure for the anisotropy in orientations of
Delaunay edges, we take the angle between an edge and the
direction of the gradient and plot the number of edges in an angular
bin (see Fig.~\ref{fig:weights}). Because we are interested in the
relative deviation from a horizontal line (which would indicate the
isotropic situation), the results are given as a fraction of the
average, $f(\theta)$. The data displayed in the topmost panel corresponds to the
 weighting schemes described above which can be thought of
as a basic correction plus a refinement thereof (the `centre of gravity' correction). 
We show the effects on the angular distribution of the edges as we apply these corrections cumulatively. 

The initial anisotropy apparent from the inclination of the line
labeled `Uncorrected' is reduced significantly (to about half its original value) by application of the Voronoi weighting alone (dashed line).  After application of the  `centre of gravity' correction, the scatter around the isotropic value of unity is below $0.25$\% except at the outer edges where the normalisation blows up the errors.

Apart from the anisotropy due to the gradient, the triangulation shows
some noise of order 0.5\% itself (the light grey area around the uncorrected lines in the three panels have a width of one standard deviation). Moreover, after all weighting has been applied, some noise remains (a unit-standard deviation area is shaded in dark grey). This noise is related to features of the triangulation itself. If the triangulation itself has more edges in a certain direction,  this cannot be corrected with a \emph{local} weighting scheme as it is a \emph{global} property subject to chance.  Such noise can be reduced by either placing more points, or, equivalently, constructing several
instances of the same triangulation and averaging over the results.

Because all the information needed for the Voronoi weighting scheme is
inherent in the tessellation and its triangulation, the computational
overhead associated with it is potentially small. Unfortunately, in most tessellation software, the areas of the Voronoi walls are not computed with the other properties of the tessellation. Calculating these areas is costly as in three dimensions the walls are generally irregular polygons with $M$ vertices where $M\geq3$. 

\subsubsection{Icosahedron weighting}

The second method is based on an `independent' division
of the unit-sphere into $M$ (approximately) equal parts. We take the
$M$ vectors (originating from the nucleus under consideration) that
point to these parts and assign weights $w_i$ to the outgoing Delaunay
edges as follows. We take a vector and calculate the $N$ dot products
with the Delaunay edges. The Delaunay edge that has the smallest dot
product gets a fraction $1/M$ added to its weight.

We have chosen the icosahedron as the basis for our weighting scheme.
We use vectors to the middle of its 30 edges together with its 12
vertices, yielding 42 reference vectors.

One significant advantage of the icosahedron method is that it is independent of
the nature of the triangulation and is therefore very flexible. It
becomes more accurate as the division of space is refined ($M$ is
increased) by taking other tessellations of the unit-sphere.  The
computational cost, however, scales linearly with $M$ forcing us to
trade off accuracy with speed. Unfortunately this `independence' has a
drawback as well: the reference vectors are oriented statically in
space, introducing a systematic bias in those directions. In order to
reduce this effect, one is forced to add some random noise to the
procedure, for example by applying random rotations of the whole
icosahedron, another costly operation.

After the correction with icosahedron weights (see middle panel of Fig.~\ref{fig:weights}, dashed line), the anisotropy has decreased below the 0.5\% level. This immediately shows the strength of this method over the more noisy Voronoi scheme (also compare the unit standard deviation regions in dark grey).

\subsubsection{Distance weighting}

Empirically we have found that in 3D the square of the distance between
nuclei is also a robust estimator of $w_i$. This quantity is readily
calculated and is by far the fastest solution to the problem at hand. After the initial correction (see bottom panel of Fig.~\ref{fig:weights}, dashed line) the anisotropy of edges has been diminished to values lower than $0.5$\% even slightly better than in the Icosahedron scheme. The deviations from the ideal isotropic case after the COG correction are of the same order as in the Voronoi and the Icosahedron case.

We have not found a valid explanation for the success of this method. Intuitively, one would think that the opening angle, $\Omega$, of a Voronoi wall with respect to its nucleus would scale as $r^{-2}$ rather than $r^{2}$. In order to find a mathematical reason for the proportionality between $\Omega$ and $r^{2}$ one is forced to delve deep into the field of computational geometry, an endeavour clearly beyond the scope of this text.

\subsection{Ballistic weighting}\label{sec:weighting_scheme_ball}

Assigning weights to the outgoing edges as described above removes the drift and clustering problems for diffusive transport described in
Sec.\ref{sec:anis_its_cons}. 

The effects of anisotropy in the ballistic case are
intrinsically more challenging to correct because it is not a
priori clear how the anisotropy of all outgoing edges of a nuclues should
be used to choose three weights for the outgoing edges.

As described in Sec.\ref{sec:ball_transp_intr} and  Sec.\ref{sec:effects_due_anis}, in the ballistic case we must distinguish between decollimation and deflection of which the first one dominates the overall `loss of direction'.

In order to diminish the decollimation of a beam, we can assign weights to the most straightforward pointing edges. If we make these weights somehow proportional to the inner product of the edge and the incoming direction, the `straightest' edges get most of the photons.

There are in fact many possibilities to assign weights to the $D$ most straightforward edges, optimising different aspects of the transportation process. If the exact direction is important, for instance, the weights should be chosen such that the vector sum of the resulting edges points straight ahead. On the other hand, minimising the decollimation will in general yield a different set of weights, those that maximise the length of the vector sum. 

From this myriad of possibilities we choose a very simple and computationally cheap approach. Edge $j$ gets weight $w_j$ given by
\begin{equation}
  w_j = {{\cos(\theta_j)/\sin(\theta_j)}\over{\sum_j \cos(\theta_j)/\sin(\theta_j)}},
  \label{eq:ballWeights}
\end{equation}
where $\theta_j$ is the angle between the incoming direction and edge $j$. Division by the sine of the angle places extra emphasis on the edges with smallest inner product. 

\begin{figure}[!ht]
  \begin{center}
    \includegraphics[width=0.45\textwidth]{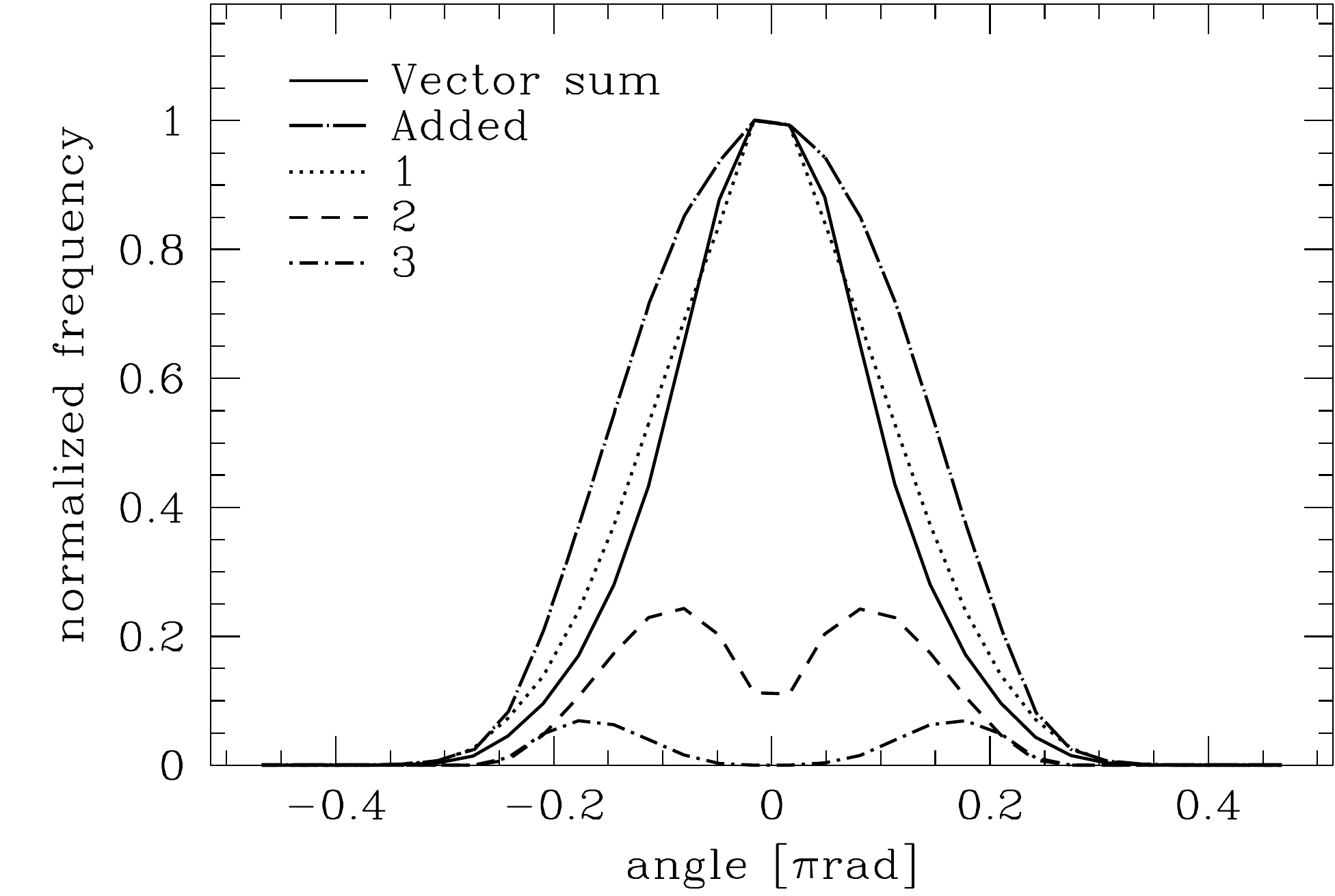}
    \caption{Normalised distribution of angles between the incoming direction and the vector sum of the (three) most straightforward pointing edges
      (solid line) and the separate most straightforward edges (dotted, dashed and dot-dashed lines) for a homogeneous distribution of nuclei with ballistic weights. The distribution of the most straightforward edges added is shown as the long dot-dashed line with label `Added'. The standard deviation (FWHM$/\sqrt{\ln 256}$) is about
     \bvalangweights for the added edges and \thetadefweights for the
      vector sum of the edges. Note that the width of the distribution of the single most straightforward edge poses a definite lower limit for the width of the `Added' lines.}\label{fig:ballweights}
  \end{center}
\end{figure}

Application of this weighting scheme on the point distribution used to produce Fig.~\ref{fig:balldev} yields the distribution of edges shown in Fig.~\ref{fig:ballweights}.  The ballistic weights as specified by Eq.(\ref{eq:ballWeights}) have the desired effect of decreasing the width of the distributions of the edges used in ballistic transport. 

In Fig.~\ref{fig:balldiff} evolution of the step-size of ballistic transport in the weighted case is shown with crosses. The decollimation angle per step is substantially smaller than in the case without weighting (\bvalangweights and \bvalang respectively). This decrease in decollimation angle consequently relaxes the requirement that the number of ballistic steps must stay below 5. Including ballistic weights typically allows for up to 12 ballistic steps before the original direction has been lost.

Lastly, there are situations where the anisotropy of edges emanating
from a cell cannot be solved by any weighting scheme. If there is no
edge in a given direction, the radiation cannot go there, whatever the
weights are. Loosely speaking, one should start off with a reasonably
isotropic triangulation in order to apply a weighting scheme in a
useful way.  In the case of Delaunay triangulations in three-space,
this is almost always the case.

\subsection{DCT throughout the grid}

A conceptually different solution for the problems described in Section~\ref{sec:anis_its_cons} is to use the direction conserving transport from Sec.~\ref{sec:DCT} throughout the simulation domain. As the angular direction of the radiation is effectively decoupled from the grid in this approach, drift, clustering, decollimation and deflection are solved for at once. The cone of nuclei that will receive radiation will however still be dependent on the grid itself as already pointed out in Sec.~\ref{sec:DCT}.

\section{Numerical examples}\label{sec:numerical_examples}

Having considered the various systematic effects inherent to the
\emph{SimpleX\/} algorithm, we now present a number of sample cases demonstrating them. In general, all of the effects described above occur simultaneously, but for transparency we analyse them in isolation.

\subsection{Effects on diffusive transport: drift}\label{sec:effects_diff}

We now proceed by describing the numerical experiment designed to show
the effects of diffuse drift. The simulation domain of unity volume in three
dimensions is filled with 5 10$^5$ nuclei subject to a gradient in the
point density of the form $n(x)= x$. We choose this linear form because it locally approximates every other type of gradient. The number of nuclei results in roughly 32 steps across the box along the gradient direction and 38 in the directions perpendicular to the gradient. These numbers allow for most of the photons to travel through the box for more than 100 steps before being captured in the absorbing boundaries.

At the site closest to the centre of the domain, a number of photons is placed. Neither the outcome nor the speed of our simulation depends on this number as we use floating point numbers to represent phtons.

Photons are transported over this grid using diffusive transport without
absorption. With every sweep the photon cloud is expected to grow in
size. In the case of a homogeneous point density the photons will be
distributed normally, as must be expected for pure diffusion.  The
gradient in the point density will distort the form of the
distribution function because of two reasons. First, the mean free
path on the grid, ${\cal D}$, scales with the point density according to
Eq.(\ref{eq:edge_length}) alowing photons to diffuse faster into the
underdense regions where the step sizes are larger. Second, the drift
phenomenon described in Sec.~\ref{sec:effects_diff_transp} will counteract this physical diffusion and will move the
cloud into the overdense region. 

In order to separate these two effects, we have performed a one dimensional Monte
Carlo simulation of 2 $10^6$ `random walkers' that take steps with a size given by 
the recipe of Eq.(\ref{eq:edge_length}). Essentially, the experiment
described above is emulated in one dimension and without the Delaunay grid as
underlying structure. The random walkers are thus expected to
experience the \emph{physical} diffusion into the underdense region only. The
unphysical drift is exclusively due to the Delaunay grid and will not
show up in the results. 

\begin{figure}[!ht]
  \begin{center}
    \includegraphics[width=0.45\textwidth]{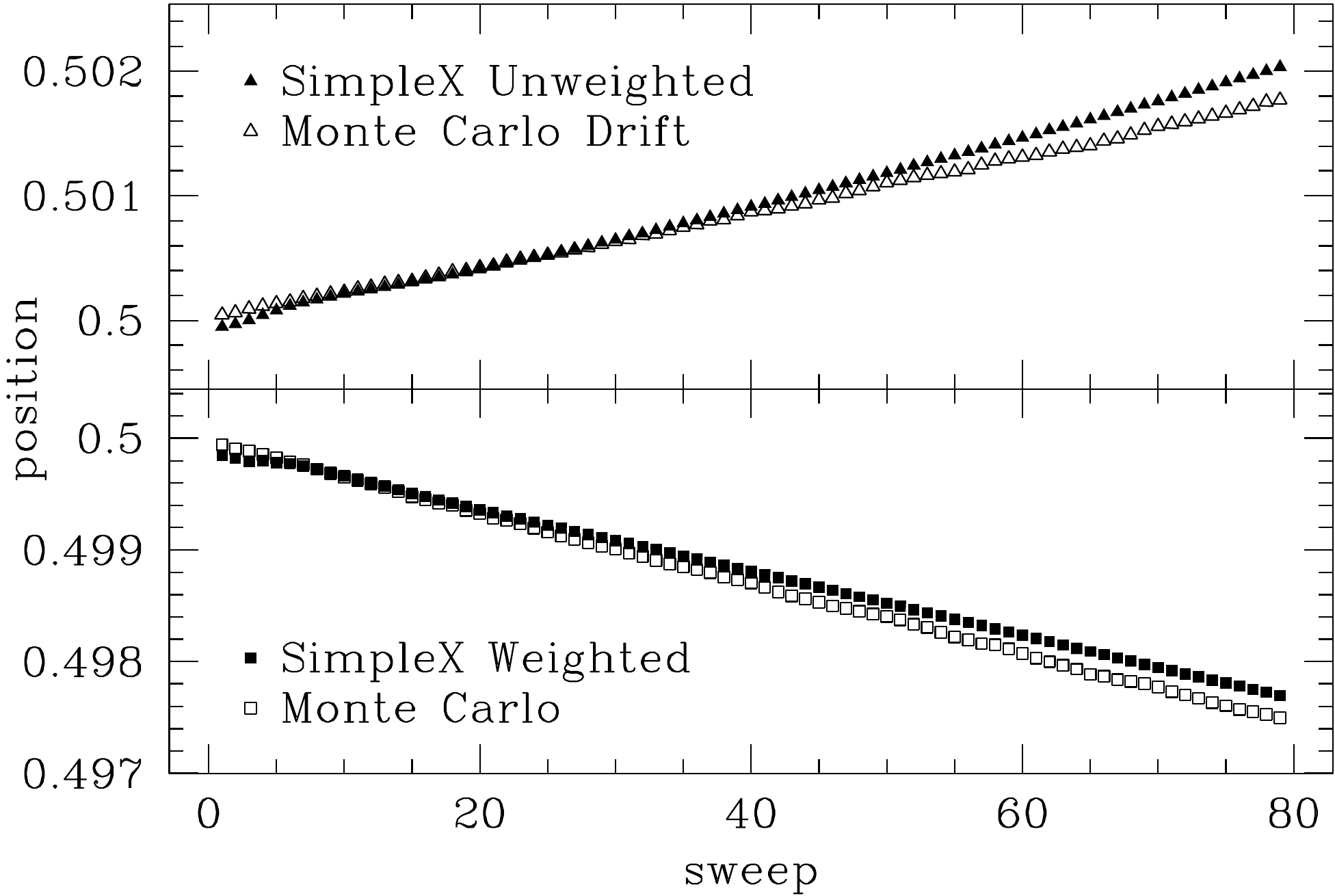}
    \caption{Intensity weighted position of a 
      photon cloud released in the centre of a simulation domain with
      a linear gradient in the number denisty of nuclei as a function of the number of sweeps.  Results obtained with \emph{SimpleX\/} are shown with filled symbols and results obtained by the Monte Carlo experiments are indicated with open symbols. We use triangles and squares in the corrected and uncorrected case respectively. }\label{fig:diffDrift}
  \end{center}
\end{figure}

In Fig.\ref{fig:diffDrift} the intensity weighted position of the
photon cloud (along the direction of the gradient) versus the number
of sweeps in the simulation is shown for \emph{SimpleX\/} and the Monte Carlo
experiments both in the case with and without weights\footnote{We have used Icosahedron weights in this case but this choice does not influence the results.}. We can see that
the behaviour conforms to our expectation. The weighting corrected \emph{SimpleX\/} result coincides with the Monte Carlo as shown in the bottom panel.

Furthermore, we have taken our drift description Eq.(\ref{eq:excess})
and applied it to the Monte Carlo experiment thus introducing a
drift toward the overdense region as would be experienced by a
photon in \emph{SimpleX\/}. The effect counteracts the physical diffusion into the underdense region resulting in a positive slope for the position of the photon cloud as a
function of sweeps (see top panel of Fig.\ref{fig:diffDrift}.) This procedure thus provides a direct quantitative check to the correctness of Eq.(\ref{eq:excess}).

The slopes of the Monte Carlo experiment agree within a thousandth of a degree with the \emph{SimpleX\/} results both for the uncorrected and weighted results. This tells us several things: First that the weighting schemes discussed in Sec.~\ref{sec:weighting_scheme} do not only correct the orientation of the edges in a statistical sense (as shown in Fig.~\ref{fig:weights}) but also allow for correct transport over these edges. This may seem a trivial statement but it could well be that although everything seems fine in a global sense, local anomalies would prevail. The transport of photons however, depends vitally on the local correctness of the weighting scheme and is thus a more stringent test. Furthermore, the recovery of the \emph{SimpleX\/} results with the Monte Carlo experiments suggests that Eq.(\ref{eq:excess}) describes the drift phenomenon accurately in a quantitative sense.

\subsection{Effects on Diffuse Transport: Clustering}\label{sec:effects_clustering}

As seen in Sec.~\ref{sec:effects_diff_transp_clust} the expansion
of a photon cloud is stalled by small (relative to the
simulation domain) scale gradients in the grid. To show this effect we have
set up the following experiment. The simulation domain is again given
by the unit cube with a large number of photons in the centre. The
5 $10^{5}$ nuclei are distributed according to the following probablity
distribution:
\begin{equation}
  n(\vec x) = 
 0.2+ 0.8\sin^2 (\omega |\vec x|)
  \label{eq:sinusoidal}
\end{equation}
where $\omega=20$. The density of nuclei thus inhibits concentric
variations with a amplitude of 0.8 on a homogeneous background of 0.2. 

We can now prepare a similar simulation with a homogeneous
distribution of nuclei where the box-size measured in units of 
$\mathcal{D}$ is the same. Comparison of the spread of the photon cloud in
the homogenous setup with the one according to Eq.(\ref{eq:sinusoidal}) will be a direct measure of the effect described by Eq.(\ref{eq:sqrtfactor_approx}).

Application of a weighting scheme for diffusive transport as described in Sec.~\ref{sec:weighting_scheme} should remove the difference between the spread of the cloud in the two cases described above. As a check we will also compare the results to the analytical result given by Eq.(\ref{eq:clustering}) and the expected spread of the photon cloud in the homogenous case (which is simply $\mathcal{D}\sqrt{N}/2$) .

\begin{figure}[!hb]
  \begin{center}
    \includegraphics[width=0.45\textwidth]{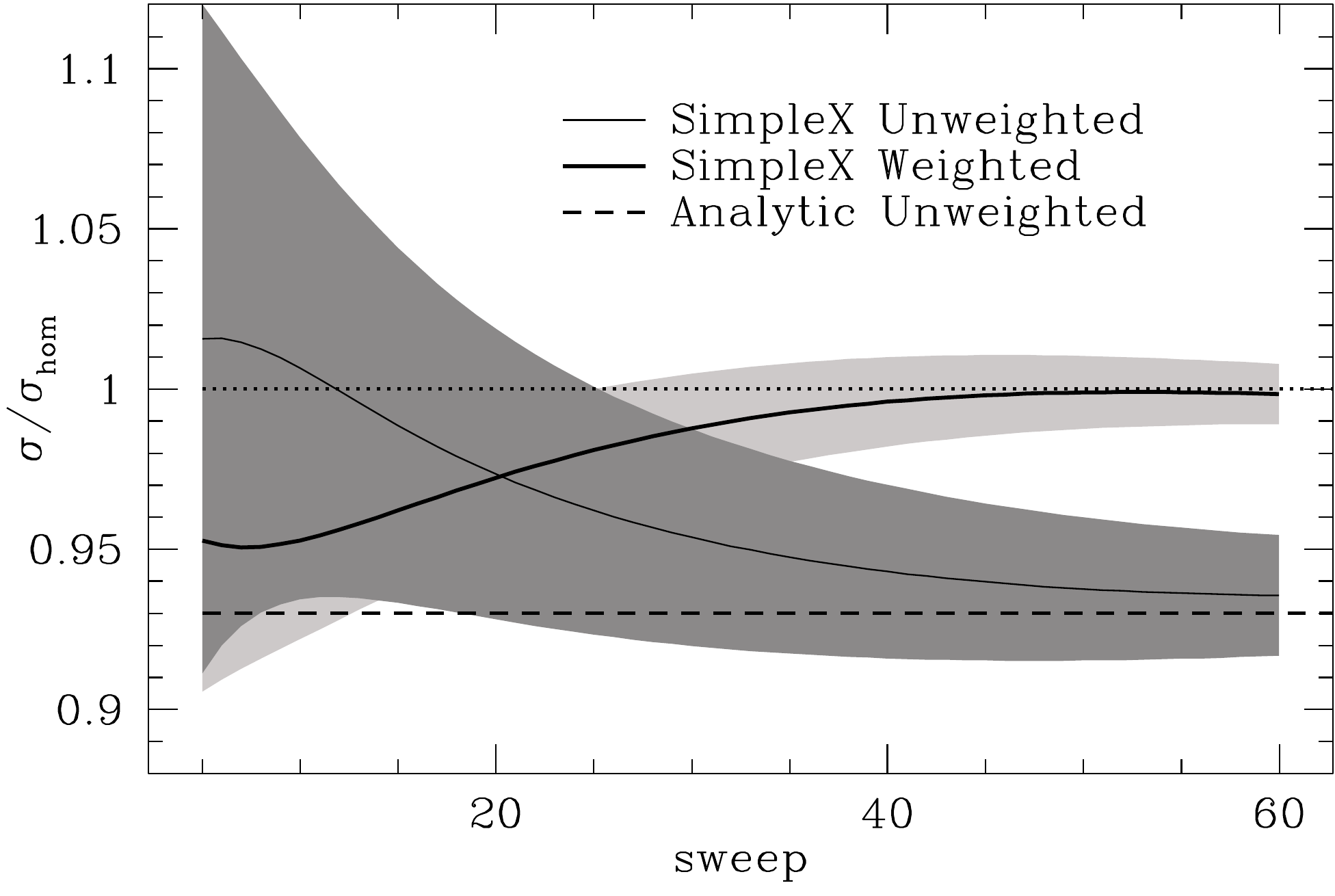}
    \caption{Normalised standard deviation of the intensity weighted positions of a photon cloud expanding in a distribution of nuclei with small scale gradients with and without weights (thick and thin solid lines respectively) and the analytical expectations according to Eq.(\ref{eq:clustering}) (dashed line). A dotted line at $\sigma/\sigma_{\mathrm{hom}}=1$ is included to guide the eye.}
   \label{fig:clustering}
     \end{center}
\end{figure}

In Fig.~\ref{fig:clustering} the spread of the photon cloud in terms of normalised standard deviation of the intensity weighted positions is shown for the inhomogeneous point density given by Eq.(\ref{eq:sinusoidal}) (solid lines) and the analytical prediction (dashed line). The data are obtained from twenty runs with different realisations of the grid. 

As expected, the photon cloud expands (almost 6\%) slower when no weights are used in the case with small scale gradients (thin solid line). Application of the weighting scheme results in a cloud that (after a slow start) conforms to the expected size (dotted line at $\sigma/\sigma_{\mathrm{hom}}=1$). The analytical result predicts this behaviour within 3\% after 30 sweeps and increasingly worse for less sweeps (although the deviation stays under 10\%). The larger deviations for fewer sweeps are due to the fact that the size of the cloud (expressed in standard deviations) is small and the results are divided by this small number. Note that the unit standard deviation region for the weighted case is less wide for the weighted case (light gray) than for the unweighted case (dark gray). This is expected as the weigting scheme corrects for local anisotropies and thus reduces noise in the diffusive transport.

We have verified that for the homogeneous case application of a weighting scheme does not alter the expansion speed of a photon cloud significantly, as expected. We can thus conclude that, although clustering can impose a substantial effect on the expansion of diffuse radiation, our weighting scheme appropriately corrects for it.

\subsection{Effects on ballistic transport: deflection}\label{sec:deflection}

In order to study if anisotropy has any effect on the direction of a beam
of photons, we set up a triangulation of a random point
distribution containing 5 $10^{5}$ nuclei placed in a cube of
unit dimension with a linear gradient along the {\it x}-direction. 
In the centre we define a spherical volume with
radius 0.05 and place a number of photons in each nucleus contained in
this volume. The photons are assigned a direction by sending them
along the edge whose direction is maximally perpendicular the {\it x}-axis. The simulation has been executed with ballistic transport using both one and three outgoing edges and additionally with the direction conserving implementation of \emph{SimpleX\/} (as described in Sec.~\ref{sec:DCT}). Every run has been repeated 10 times with different instances of the grid to suppress shot noise and obtain error estimates.

\begin{figure}[!ht]
  \begin{center}
    \includegraphics[width=0.45\textwidth]{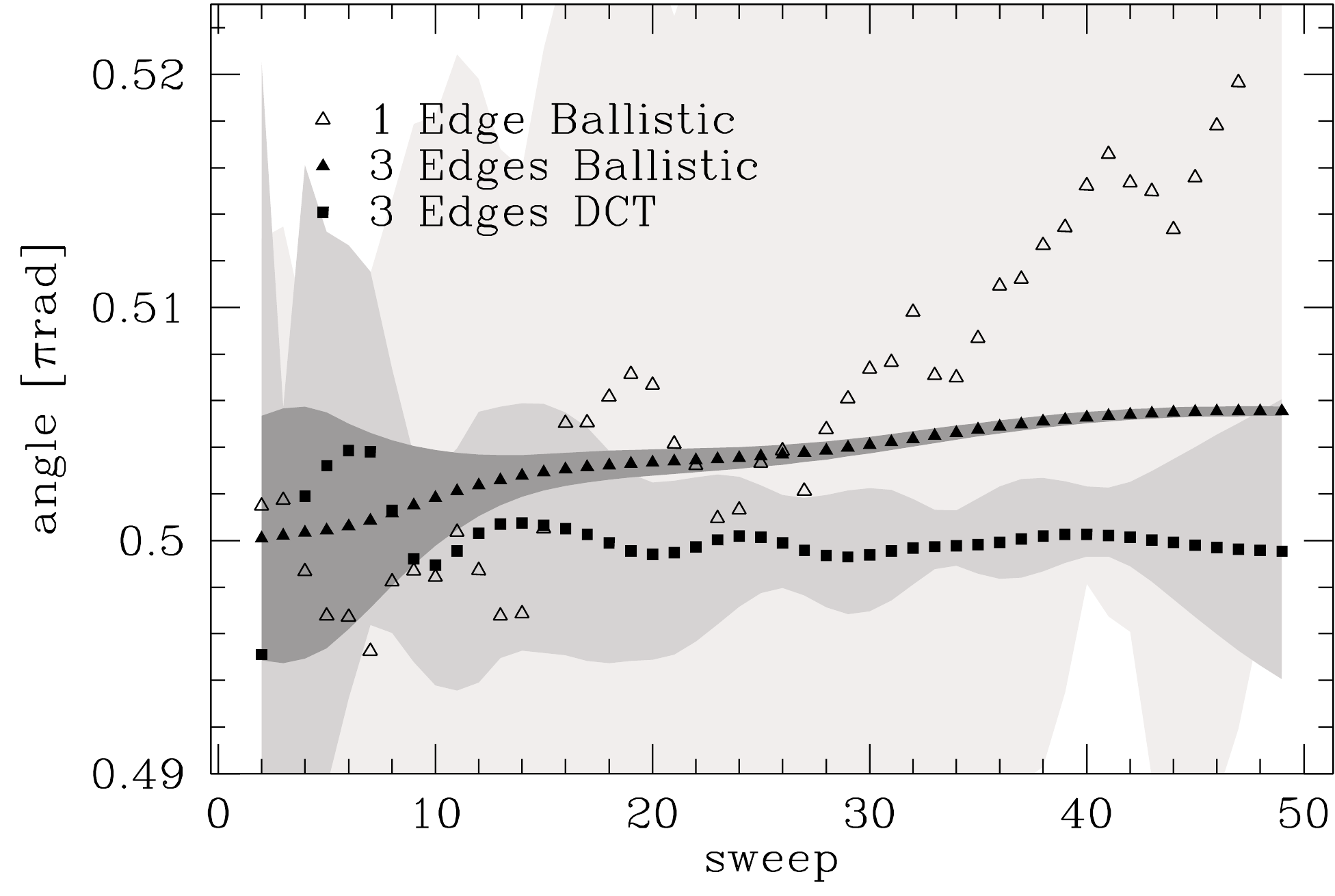}
    \caption{Mean angular directionality of a photon cloud as a function of the number of sweeps. In this simulation a linear gradient in the density of nuclei is present and the angle is measured relative to the direction of the gradient, so smaller values for the angle are pointed toward the denser region. As expected, the direction conserving implementation of \emph{SimpleX\/} (as described in Sec.~\ref{sec:DCT}, indicated by filled squares) has no long-term loss of directionality although it shows some oscillatory behaviour that dampens with time. Ballistic transport with one straightforward edge (open triangles) shows a deflection towards larger angle with number of sweeps of order $5  10^{-4}\, \pi$ rad/sweep in accordance with predictions. Using three outgoing edges (filled triangles) diminishes the effective deflection by a factor of five. Unit standard deviation regions are shaded in grey for the three simulations.}\label{fig:ballDefDynamic}
  \end{center}
\end{figure}

When applying our ballistic transport method, the photons will undergo small angular deflections into the under-dense region as predicted in sec.~\ref{sec:effects_due_anis}.
The expected effective deflection for this setup is, using Eq.(\ref{eq:theta_eff}), $\theta_{\mathrm{eff}}=5   10^{-4} \pi$ rad/sweep. Thus yielding a cumulative result of approximately $0.025 \, \pi $ rad after 50 sweeps. This estimate is nicely consistent with the result from the simulation with one straightforward edge denoted by the open triangles in Fig.~\ref{fig:ballDefDynamic}. As expected, sending radiation along three edges decreases the deflection and fluctuation of the angular direction in general because radiation is distributed over many (3$^{N_{s}}$) edges, quickly reducing shot noise. When using the direction conserving implementation of \emph{SimpleX\/} (filled squares), the radiation does not suffer from ballistic deflection. 

\section{Application}\label{sec:application}

In this section we will show how to apply the results from the analysis above to a more (astro-)physically relevant problem. In short, we ask ourselves the question: given a density (or opacity) distribution, how do we construct a Voronoi-Delaunay structure that keeps the described problems acceptably small while representing the physical problem as accurately as possible. This will often mean maximising the dynamic range with the prerequisite that low density regions still have enough resolution.

\subsection{Sampling function}

The trade-off is thus between maximising dynamic range and minimising the strong gradients this inevitably introduces. In other words: we must investigate the effects that different sampling
functions have on the triangulation and its transport properties. These functions
translate the physical medium (expressed as an opacity or density map)
to a point distribution $n(\vec x)$ whose points form the nuclei of
the triangulation. As mentioned in Sec.\ref{sec:effects_diff_transp} the
average Delaunay edge length can be associated with the mean free path
of the photons in a very natural way. This is achieved when $n(\vec
x)$ scales with the opacity (or density) to the $D$-th power
\citep[][Eq.(14)]{2006PhRvE..74b6704R}
\begin{equation}
  n(\vec x) = 
  \Phi \ast \rho^D (\vec x)
  \label{eq:point_density}
\end{equation}
where $\Phi$ denotes a homogeneous Poisson process, $D$ is the
dimension of the propagation space, and $\ast$ denotes the convolution.

One drawback of using Eq.(\ref{eq:point_density}) is that the
resulting point process, given a substantial density contrast, places
many nuclei in dense regions and very few in underdense regions.  For
many applications, simulating the true mean free path may thus lead to
a required number of nuclei that is prohibitively large. 

We therefore adopt a more flexible sampling function which behaves differently for the extremal densities of the grid. According to this function, the number density of nuclei that generate the Voronoi grid is given by
\begin{equation}
n(\vec x) = 2 \Phi \ast \left(  y^{-\alpha}  + y^{-D} \right) ^{-1}
\label{eq:sampling}
\end{equation}
where $y \equiv \rho/\rho_{0}$ and $\rho_{0}$ is a reference density that marks the transition between the two regimes of sampling power. In our applications, $\alpha$ will be always smaller than $D$ (and even smaller than unity) in order to prohibit the over-emphasis on high density regions.

According to Eq.(\ref{eq:equality_length}) and Eq.(\ref{eq:L_def}), the quantity that links the properties of the grid (density of nuclei and gradients therein) to the systematic effects described in the previous sections is $Q_{n} \equiv n/\nabla n$. We are now in the position to define, analogous to $Q_{n}$, the quantity $Q_{y} \equiv y/\nabla y$ which can be measured over the physical density (opacity) field. As the sampling function of Eq.(\ref{eq:sampling}) translates the physical field to the density of nuclei, the measured value of $Q_{y}$ and the upper limit of $Q_{n}$ (posed by the maximally acceptable value for either $L_{eq}$ or $L_{def}$) will constrain the sampling parameters $\alpha$ and $\rho_{0}$.

\subsection{Cosmological density field}

Let us now apply this idea to a more realistic example and see what this all means in practice. A typical application wherein the strength of the \emph{SimpleX\/} algorithm can be fully deployed is the epoch of re-ionisation (see \citet{JP} for an account of relevant test cases). Both the wide range in densities (four to five orders of magnitude), and the large number of sources (hundreds to thousands in realistic cosmological volumes) are treated naturally by \emph{SimpleX\/} due to the adaptive nature of the Delaunay grid and the property that computational effort does not increase with increasing number of sources.

\begin{figure*}[!ht]
  \begin{center}
    \includegraphics[width=\textwidth]{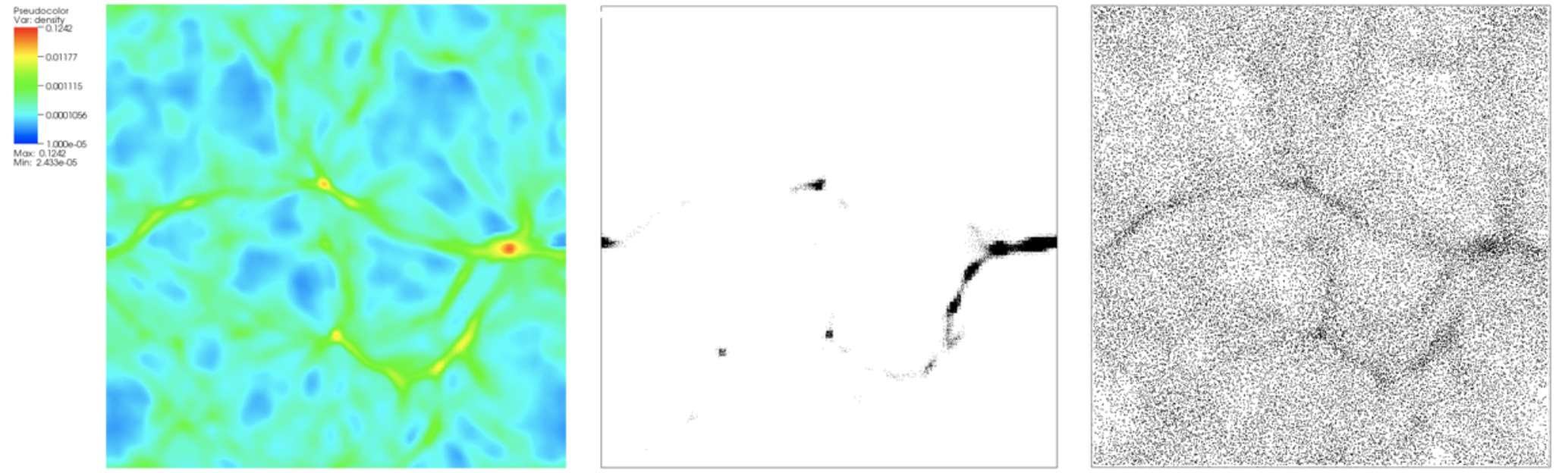}
    \caption{Cut through a cosmological density field (left panel; colours indicate number density in logarithmic scale) sampled with $10^{5}$ points using the sampling function of Eq.~\ref{eq:point_density} (middle panel) and Eq.~\ref{eq:sampling} (right panel) with the parameters obtained from analysis of the underlying density field.}\label{fig:sampling}
  \end{center}
\end{figure*}

We will now take a relatively small (500/h Mpc comoving) cosmological volume (as used in Test 4 of \cite{2006MNRAS.371.1057I}) and construct a representative point set using constraints from analysis of the original data. In this specific case the cosmological density field was presented as a regular grid of 128$^{3}$ equal size cells. 

\subsection{Constraints}

Given the regular grid, the quantity $Q_{y} $ can be determined for every pair of adjacent cells. The thus obtained mean value for $Q_{y} $ is $1.6$. 

This value implies that if we would use a linear translation between the mass density of the original data and the number density of the nuclei of the Delaunay grid, the equality length (see Eq.(\ref{eq:equality_length})) would become $8 \cdot 1.6=13$ and the deflection length of Eq.(\ref{eq:L_def}) becomes $50  \cdot 1.6=81$. This means that radiation can travel 13  box-lengths before diffuse drift starts to dominate the transport of radiation, and 81 box-lengths before ballistic deflection reaches a cumulative magnitude of $\pi/4$. 

Such a linear sampling could be performed using $\alpha = 1$ in Eq.(\ref{eq:nuclei_mass_density}) or, equivalently, $\alpha = 1$ and $ \lim_{\rho_{0} \rightarrow 0} $ in Eq.(\ref{eq:sampling}). Linear sampling is hardly ever desirable in a cosmological setting, however, as it tends to place the bulk of the nuclei in high density regions resulting in over-sampling (many nuclei per grid cell of the original grid) of filaments and clumps and under-sampling of voids. 

Let us now proceed by finding optimal values for our sample parameters, given the measured value of $Q_{y}$ and a chosen value for the maximal allowed equality length.  An upper limit for the equality length, through Eq.(\ref{eq:equality_length}), immediately implies a maximal $Q_{n}$.

If, for example, we require $L_{eq}=40$ (and consequently $L_{\mathrm{def}}=250$) we see that we need to choose $\alpha$ and $\rho_{0}$ such that $Q_{n}=5$. From Eq.(\ref{eq:sampling}) we find the expression that links the measured value for $Q_{y}$ to $Q_{n}$ and thus to the sample parameters
\begin{equation}
Q_{n} =
	Q_{y}
	\frac{ y^{-\alpha}  + y^{-D} }{ y \left(  \alpha y^{-\alpha-1}  +D y^{-D-1} \right) }.
\label{eq:nOverGrad}
\end{equation}
Solutions for $Q_{n} =Q_{y} $ exist only for certain pairs of $\alpha$ and $\rho_{0}$ as can be seen from the limit of Eq.(\ref{eq:nOverGrad})
\begin{equation}
Q_{n}^{lim}= \lim_{y \rightarrow \infty} Q_{n} = \frac{Q_{y}}{\alpha}
\label{eq:Qlim}
\end{equation}
So, if we would choose $\alpha$ at its maximal value of $Q_{y}/Q_{n}^{lim}$, we would effectively force $\rho_{0}\rightarrow 0$, resulting in the limit where all density regimes are sampled with the same power, $\alpha$.
Choosing $\alpha$ to be smaller but close to its extremal value maximises the dynamic range in the high density regions and allows $\rho_{0}$ to attain values where still a substantial volume of the density field (namely the voids) is sampled with the advantageous power $D$.

Solving Eq.(\ref{eq:Qlim}) for  $Q_{y} =1.6$ we find $\alpha = 0.32$ as the maximal value. Taking the somewhat smaller value of 0.3, the corresponding value for $\rho_{0}$ is $3.7\times 10^{-5}$ cm$^{-3}$ (well larger than the smallest densities in the cosmological field which are about $2.4 \times 10^{-5}$). 

\subsection{The grid}

In the right panel of Fig.~\ref{fig:sampling}  a cut at $z=0.5$ through the sampling obtained with these parameters is shown. The middle panel shows the point distribution one gets when sampling with $\alpha=3$ throughout the grid. Evidently, this distribution does little justice to the low-density regions while placing almost all points in the dense filaments and their intersections whereas the `hybrid' sampling method places sufficient resolution in both the low- and high-density regions. Moreover, for the  $\alpha=3$ sampling, the cell size in the filaments and dense clumps is many times smaller than the that of the original regular grid. No information is carried by this extra resolution, however, effectively wasting computational resources. As an extra constraint (apart from 
$L_{eq}$ and $L_{\mathrm{def}}$) one could demand that the smallest Voronoi volume is equal to the resolution of the original data. This could either be the highest AMR resolution or, in the case of SPH, the minimal smoothing length.

\subsection{Speed considerations}

Although we will not delve into the details here, the speed of  \emph{SimpleX\/} also depends on the form of the grid. Because in every sweep all nuclei are visited exactly once, the execution time of the method scales linearly with the number of nuclei. In turn, the number of sweeps needed to obtain a converged roughly scales as the number of sweeps it takes to enable every point within the simulation domain to `communicate' its physical properties with every other other location by exchange of photons.

Now consider the point distribution in the middle panel of Fig.~\ref{fig:sampling}. The resulting grid will have many short edges in the high density clump in the east of the domain. For diffusive transport or ballistic transport over many (more than roughly 5) edges such a clump will trap photons for many sweeps as their travelled distance after $N$ sweeps is given by $\sqrt{N} \mathcal{D}$ where $\mathcal{D}$ is small. This is another reason against the use of high sampling powers throughout the simulation domain.

\section{Discussion}\label{sec:discussion}

Now that we have identified and corrected the various effects due to anisotropy in the Delaunay grid we can take a step back and discuss their implications in a broader context.  

\subsection{Prevention versus correction}

It is impossible to define the perfect grid in general. As already pointed out, a large dynamic range inevitably leads to gradients that, in turn, give rise to the unphysical effects described in this text. Although we showed that these effects can be corrected for by weighting schemes, these corrections are not fail-safe if the gradients are too strong. In particular, when the density of nuclei changes appreciatively on length scales smaller than $\mathcal{D}$ (or, more precise, $Q_{n} \leq 1$), the lack of edges towards the under-dense region may lead to extreme decollimation as there simply are hardly any edges that point that way. In this case, a weighting scheme is not feasible as there are no edges to assign weights to in the under-dense direction. Simply said, we should start with a fairly well behaved grid in the first place. 

One could argue that if the grid is constructed such that the unphysical effects are below a predefined tolerance, the deployment of weighting schemes can be circumvented. On the other hand, this puts rather stringent constraints on the grid and it might be preferable to retain more dynamic range (and thus structure) in the grid at the expense of the need for weighting.

An intermediated solution will often be the most viable option. Retaining the dynamic range of the data as much as possible while weighting schemes and DCT are employed at \emph{certain locations} in the simulation domain only.

\subsection{Number of steps versus $L_{eq}$ and $L_{\mathrm{def}}$}

In Sec.~\ref{sec:anis_its_cons} we have derived typical length scales at which 
the unphysical effects become important in a relative sense. Being cautious, we have set these length scales to many times the size of the simulation domain but this is not necessary in most real life situations. This is because radiation typically travels only moderate distances along the grid. It might therefore be more intuitive to think about the number of steps a photon travels along the grid before being scattered or absorbed instead of $L_{eq}$ and $L_{\mathrm{def}}$.  In either the event of absorption or scattering, the past of the photon is essentially erased and it starts a new life, so decollimation, deflection and drift are basically `reset' to zero. 

We can thus assess the possible harm of anisotropy in a simulation by monitoring the fraction of Delaunay length and physical mean free path, $R$, defined as
\begin{equation}
R \equiv \frac{\mathcal{D}}{l_{\mathrm{mfp}}}
\end{equation}

First of all, $R$ should not be much smaller than 1, because this
would result in decollimation and in extreme cases even in
simulating straight light rays by a random walk (see also the discussion in Sec.~\ref{sec:ball_transp_intr}). 

In the limit of large $R$, however, a large portion of the photons will be absorbed after
travelling along this edge. This is merely a resolution issue; the
photons will never travel further than one Voronoi region away from
the place where they ought to have been absorbed. 

Using a sampling function of the form of Eq.(\ref{eq:point_density}) would result in $R$ being a constant along the grid. If we use a more general sampling method, $R$ is no longer a global constant. We can however still adjust the local optical depth for absorption with the local value of $R$. If we do so, the place where a photon finally gets
absorbed or leaves the simulation domain still has the right
probability distribution, but the interpretation of intermediate
photon densities in the simulation becomes less intuitive. Already the
photon densities are no `time' densities but `step' densities, but now
photons in regions with different densities will travel a different
amount of mean free paths per sweep. Still, if we are interested in
the effect of the photons on the medium and in the directions of
photons that leave the box, we are justified to sample with any other power than 3, as long as we make local corrections to the optical depth proportional to $R$.

\subsection{A dynamic grid}

We have argued that the ratio between the local mean free path and the Delaunay edges should be chosen with care in order to ensure meaningful results. Because the local mean free path can change due to changes in the medium properties, the Delaunay
edge length should adapt accordingly to keep $R$ fixed.

Recently the \emph{SimpleX\/} code has been extended to handle dynamic media \citep{JP}, meaning that nuclei can be added or deleted during the simulation. This allows
the triangulation to adjust itself to the (changing) local properties
of the medium. 

We can set up strict conditions under which a nucleus
gets deleted (and its attributes divided between its neighbours). A possible condition could be: delete a nucleus whenever the local mean free path of the ionising photons exceeds the average length of its outgoing edges. In this way we keep deflection and decollimation acceptably small in the case of ballistic transport. 

\section{Future work}

In this section we discuss several limitations of the current \emph{SimpleX\/} method and indicate a possible route to overcome these limitation in future versions of the code.

\subsection{Multiple frequencies}

In the current incarnation of \emph{SimpleX\/}, a grey approximation is used where the effect of a source spectrum is taken into account by the use of effective, intensity weighted, cross sections. 
Implementation of a full multi-frequency treatment is not trivial as it would formally require a separate grid for every new frequency bin. In practice, however, we have found that by transporting similar frequencies over the same grid, the computational demands will not become prohibitive. A version of \emph{SimpleX\/} incorporating multiple frequencies is currently under development and we postpone a discussion of related issues to future articles. 

\subsection{Cosmological redshift}

For cosmological simulations, the effect of expanding space will result in shifting of frequency of the transported photons. In our current one-frequency approach this is nothing more than a gradual loss of ionizing energy for the photons. In future multi-frequency versions of the code more care should be taken of this issue especially when line-transfer is included.

In principle, also relativistic `beaming' could play a non-neglegible role in simulations of large cosmological volumes. Due to expansion of the universe, an absorbing parcel of gas will have different relative velocities with sources at different distances. The resulting enhancement of the observed intensity \citep[e.g. Eq. 4.9 in ][]{1986rpa..book.....R} can be substantial. Such beaming could in principle be modeled in \emph{SimpleX\/} by adjusting the weighting factors for diffuse and ballistic transport appropriately but we leave a detailed analysis for future work.

\subsection{Strong scattering}

Finally, in the case of strong scattering (if the cross section for absorption is negligibly small compared to the cross section for scattering) the \emph{SimpleX\/} method only converges asymptotically to the correct answer. Colloquially speaking this is due to the fact that transport happens on a cell-to-cell basis and this can take a large number of sweeps to bring all scattering particles in the simulation domain into equilibrium.
This problem is well known and has been overcome by classical radiative transfer methods by the use of so-called accelerated Lambda iteration methods. 

In the cosmological applications of \emph{SimpleX\/}, strong scattering is not a dominant process, however, and we will not pursue an implementation of accelerated Lambda iteration schemes in our method in the near future.

\section{Summary}\label{sec:summary}

\begin{enumerate}
\item The \emph{SimpleX\/} algorithm is a relatively new and non-standard method and as such deserves a thorough analysis of its systematic effects. This assessment has two important purposes: First, exploring the boundaries of its reliability and accuracy and, second, increasing its region of application through improvements.
\item The mathematically transparent nature of the \emph{SimpleX\/} algorithm allows for a rigorous and general assessment of its systematic effects. Although only some of the described effects (diffuse drift and ballistic decollimation) need to be considered in typical applications \citep[see][]{JP}, we have investigated all four in detail for completeness.
\item The use of a random Delaunay triangulation in the radiative transfer
  method \emph{SimpleX\/} introduces global errors when the point-density is
  inhomogeneous. We have identified and quantified four distinct
  effects: `drift' and `clustering' of photons in diffusive transport, and `decollimation' and `deflection' in ballistic transport.

\item We show that decollimation become troublesome only when the number of traversed edges becomes larger than roughly 5 steps. This implies that one either preclude this regime or, when this is undesirable or impossible, correct for the unphysical behaviour.
\item A weighting method for ballistic transport has been shown to decrease the effect of decollimation significantly pushing the limit for correct transport from 5 to 12 steps.
\item We have shown how drift and clustering can be adequately corrected for by
  adopting a weighting scheme. Three weighting schemes for diffusive transport have been discussed and compared.
\item The use of a direction conserving variant of the ballistic transport has been introduced as an alternative to the weighting procedure and to prevent loss of angular direction in optically thin regions. This approach provides a means of correcting for deflection, drift and clustering as well. The computational cost and need for additional randomisation of the solid angle directions makes the employment of DCT not generally favourable, however.
\item Finally, we have applied our analysis to an astrophysically relevant example. This elucidates the intimate relation between the construction of a computational grid and the possible need for weighting schemes.

\end{enumerate}
\begin{acknowledgements}
      Part of this work was supported by the German
      \emph{Deut\-sche For\-schungs\-ge\-mein\-schaft, DFG\/} project
      number Ts~17/2--1.
      CJHK would like to extend his gratitude to Rien van de Weijgaert, Jakob van Bethlehem, Sven de Man en Marcel Kruip for proofreading the manuscript and valuable discussion. We also thank the anonymous referee for useful comments on the manuscript.
\end{acknowledgements}

\bibliographystyle{aa}

\begin{thebibliography}{}

\bibitem[{Abbott \& Lucy(1985)}]{Abbott:1985p1320}
Abbott, D.~C. \& Lucy, L.~B. 1985, Astrophysical Journal, 288, 679

\bibitem[{{Delone}(1934)}]{Delone1934}
{Delone}, B.~N. 1934, Bull. Acad. Sci. USSR: Classe Sci. Mat, 793, 7

\bibitem[{Du {et~al.}(1999)Du, Faber, \& Gunzburger}]{du:637}
Du, Q., Faber, V., \& Gunzburger, M. 1999, SIAM Review, 41, 637

\bibitem[{{Edelsbrunner}(2006)}]{edelsbrunner}
{Edelsbrunner}, H. 2006, Geometry and topology for mesh generation, 1st edn.
  (Cambridge University Press)

\bibitem[{{Icke} \& {van de Weygaert}(1987)}]{1987A&A...184...16I}
{Icke}, V. \& {van de Weygaert}, R. 1987, \aap, 184, 16

\bibitem[{{Iliev} {et~al.}(2006){Iliev}, {Ciardi}, {Alvarez}, {Maselli},
  {Ferrara}, {Gnedin}, {Mellema}, {Nakamoto}, {Norman}, {Razoumov},
  {Rijkhorst}, {Ritzerveld}, {Shapiro}, {Susa}, {Umemura}, \&
  {Whalen}}]{2006MNRAS.371.1057I}
{Iliev}, I.~T., {Ciardi}, B., {Alvarez}, M.~A., {et~al.} 2006, \mnras, 371,
  1057

\bibitem[{{Lloyd}(1982)}]{Lloyd1982}
{Lloyd}, S.~P. 1982, IEEE TRANSACTIONS ON INFORMATION THEORY, IT-28, no. 2, 129

\bibitem[{{Mellema} {et~al.}(2006){Mellema}, {Iliev}, {Alvarez}, \&
  {Shapiro}}]{2006NewA...11..374M}
{Mellema}, G., {Iliev}, I.~T., {Alvarez}, M.~A., \& {Shapiro}, P.~R. 2006, New
  Astronomy, 11, 374

\bibitem[{{Mihalas} \& {Weibel Mihalas}(1984)}]{1984frh..book.....M}
{Mihalas}, D. \& {Weibel Mihalas}, B. 1984, {Foundations of radiation
  hydrodynamics} (New York: Oxford University Press)

\bibitem[{{Okabe} {et~al.}(2000){Okabe}, {Boots}, {Sugihara}, {Chiu}, \&
  {Kendall}}]{okabe}
{Okabe}, A., {Boots}, B., {Sugihara}, K., {Chiu}, S.~N., \& {Kendall}, D.~G.
  2000, Spatial tessellations: concepts and applications of Voronoi diagrams,
  2nd edn. (Wiley)

\bibitem[{{O'Rourke}(2001)}]{orourke}
{O'Rourke}, J. 2001, Computational Geometry in C, 2nd edn. (Cambridge
  University Press)

\bibitem[{{Paardekooper} {et~al.}(2009){Paardekooper}, {Kruip}, \& {Icke}}]{JP}
{Paardekooper}, J.-P., {Kruip}, C. J.~H., \& {Icke}, V. 2009, to be submitted.

\bibitem[{{Ritzerveld} \& {Icke}(2006)}]{2006PhRvE..74b6704R}
{Ritzerveld}, J. \& {Icke}, V. 2006, \pre, 74, 026704

\bibitem[{{Ritzerveld}(2007)}]{2007PhDT1R}
{Ritzerveld}, N.~G.~H. 2007, PhD thesis, Leiden Observatory, Leiden University,
  P.O.~Box 9513, 2300 RA Leiden, The Netherlands

\bibitem[{{Rybicki} \& {Lightman}(1986)}]{1986rpa..book.....R}
{Rybicki}, G.~B. \& {Lightman}, A.~P. 1986, {Radiative Processes in
  Astrophysics}, ed. {Rybicki, G.~B.~\& Lightman, A.~P.}

\bibitem[{Schaap \& {Van de Weygaert}(2000)}]{Schaap:2000p855}
Schaap, W.~E. \& {Van de Weygaert}, R. 2000, Astronomy and Astrophysics, 363,
  L29

\bibitem[{{Vedel Jensen}(1998)}]{vedeljensen}
{Vedel Jensen}, E.~B. 1998, Local Stereology, 1st edn. (World Scientific)

\bibitem[{{Voronoi}(1908)}]{Voronoi1908}
{Voronoi}, G. 1908, J. Reine Angew. Math., 134, 198

\end{thebibliography}

\end{document}